\input harvmac.tex
\input amssym.tex

\def\newdate{22/12/2004}

\def\a{\alpha}
\def\b{\beta}
\def\g{\gamma}
\def\l{\lambda}
\def\d{\delta}
\def\e{\epsilon}
\def\t{\theta}

\def\s{\sigma}

\def\G{\Gamma}
\def\L{\Lambda}

\def\p{\partial}
\def\half{{1\over 2}}

\def\CL{{\cal L}}
\def\CM {{\cal M}}
\def\CD{{\cal D}}
\def\CDD{\tilde{\cal D}}

\def\ie{{\it i.e.\ }}

\Title{\vbox{
\hbox{CERN-PH-TH/2004-205}
\hbox{YITP-SB-04-66}
}}
{\vbox{\centerline{Super-Chern--Simons Theory 
as Superstring Theory}
}}
\smallskip
\medskip\centerline{
{\bf Pietro~Antonio~Grassi}$^{~a,b,c,}$\foot{pgrassi@cern.ch}
and 
{\bf Giuseppe~Policastro}$^{~d,}$\foot{policast@theorie.physik.uni-muenchen.de}} 
\medskip 
\centerline{\it $^{a}$ CERN, Theory Division, CH-1121 Genev\'e 23, Switzerland,}  
\centerline{\it $^{b}$ YITP, SUNY, Stony Brook, NY 11794-3840, USA,} 
\centerline{\it $^{c}$ DISTA, Universit\`a del Piemonte Orientale,}
\centerline{\it ~~~~ Via Bellini, 1 15100 Alessandria, Italy}
\vskip .3cm
\centerline{$^{d}$ {\it Ludwig-Maximilians-Universit\"at,
Theoretische Physik} }     
\centerline{\it Theresienstr. 37, 80333 M\"unchen, Germany}

\bigskip

\noindent

Superstrings and topological strings with supermanifolds as 
target space play a central role in the recent developments in string theory. 
Nevertheless the rules for higher-genus computations are still 
unclear or guessed in analogy with bosonic and fermionic strings. 
Here we present a common geometrical setting to develop 
systematically the prescription for amplitude computations. The 
geometrical origin of these difficulties is the theory of integration 
of superforms. We provide a translation between the theory of supermanifolds 
and topological strings with supertarget space. We show how in this
formulation one can naturally construct   
picture changing operators to be inserted in the correlation functions
 to soak up the zero modes of 
commuting ghost and we derive the amplitude prescriptions from the
coupling with an extended topological gravity on the worldsheet. As an
application we consider a simple model   
on ${\bf R}^{(3|2)}$ leading to super-Chern-Simons theory. 

\Date{\newdate}

\lref\as{A.~Schwarz, 
  Commun.Math.Phys. 155 (1993) 249-260;
M. Alexandrov, M. Kontsevich, A. Schwarz, O. Zaboronsky,
      Int.J.Mod.Phys. A12 (1997) 1405-1430;
        A.S.~Cattaneo, G.~Felder
   math.QA/0102108.}

\lref\BerkovitsFE{ 
N.~Berkovits, 
JHEP { 0004}, 018 (2000) 
[hep-th/0001035]. 
} 

\lref\BerkovitsPH{ 
N.~Berkovits and B.~C.~Vallilo, 
JHEP { 0007}, 015 (2000) 
[hep-th/0004171]. 
} 

\lref\GrassiUG{
P.~A.~Grassi, G.~Policastro and P.~van Nieuwenhuizen,
JHEP {\bf 0211}, 001 (2002)
[arXiv:hep-th/0202123];

P.~A.~Grassi, G.~Policastro, M.~Porrati and P.~Van Nieuwenhuizen,
JHEP {\bf 0210}, 054 (2002)
[arXiv:hep-th/0112162];

P.~A.~Grassi, G.~Policastro and P.~van Nieuwenhuizen,
Adv.\ Theor.\ Math.\ Phys.\  {\bf 7}, 499 (2003)
[arXiv:hep-th/0206216].
}

\lref\GrassiXF{
P.~A.~Grassi, G.~Policastro and P.~van Nieuwenhuizen,
Phys.\ Lett.\ B {\bf 553}, 96 (2003)
[arXiv:hep-th/0209026].
}

\lref\WittenNN{
E.~Witten,
arXiv:hep-th/0312171.
}

\lref\BerkovitsHG{
N.~Berkovits,
Phys.\ Rev.\ Lett.\  {\bf 93}, 011601 (2004)
[arXiv:hep-th/0402045].
}

\lref\NeitzkePF{
A.~Neitzke and C.~Vafa,
arXiv:hep-th/0402128.
}

\lref\BerkovitsUE{
N.~Berkovits and P.~S.~Howe,
Nucl.\ Phys.\ B {\bf 635}, 75 (2002)
[arXiv:hep-th/0112160].
}

\lref\BerkovitsYR{
N.~Berkovits and O.~Chandia,
Nucl.\ Phys.\ B {\bf 596}, 185 (2001)
[arXiv:hep-th/0009168].
}

\lref\GreenZG{
M.~B.~Green and J.~H.~Schwarz,
Nucl.\ Phys.\ B {\bf 181}, 502 (1981);
M.~B.~Green and J.~H.~Schwarz,
Nucl.\ Phys.\ B {\bf 198}, 252 (1982);
M.~B.~Green and J.~H.~Schwarz,
Nucl.\ Phys.\ B {\bf 198}, 441 (1982).
}

\lref\SethiCH{
S.~Sethi,
Nucl.\ Phys.\ B {\bf 430}, 31 (1994)
[arXiv:hep-th/9404186].
}

\lref\AganagicYH{
M.~Aganagic and C.~Vafa,
arXiv:hep-th/0403192.
}

\lref\KumarDJ{
S.~P.~Kumar and G.~Policastro,
arXiv:hep-th/0405236.
}

\lref\SchwarzAK{
A.~Schwarz,
Lett.\ Math.\ Phys.\  {\bf 38}, 91 (1996)
[arXiv:hep-th/9506070];
A.~Konechny and A.~Schwarz,
arXiv:hep-th/9706003;
A.~A.~Voronov, A.~A.~Roslyi and A.~S.~Schwarz,
Commun.\ Math.\ Phys.\  {\bf 120}, 437 (1989).
}

\lref\HoriKT{
K.~Hori and C.~Vafa,
arXiv:hep-th/0002222.
}

 \lref\BookTop{Kentaro Hori et al.
{\sl Mirror Symmetry}, Clay Mathematics Monographs, V. 1, 2003}

\lref\BerkovitsPX{
N.~Berkovits,
JHEP {\bf 0409}, 047 (2004)
[arXiv:hep-th/0406055].
}

\lref\AnguelovaPG{
L.~Anguelova, P.~A.~Grassi and P.~Vanhove,
arXiv:hep-th/0408171.
}

\lref\FriedanGE{
D.~Friedan, E.~J.~Martinec and S.~H.~Shenker,
Nucl.\ Phys.\ B {\bf 271}, 93 (1986).
}

\lref\VerlindeSD{
E.~Verlinde and H.~Verlinde,
Phys.\ Lett.\ B {\bf 192}, 95 (1987).
}

\lref\KnizhnikAK{
V.~G.~Knizhnik,
Sov.\ Phys.\ Usp.\  {\bf 32}, 945 (1989)
[Usp.\ Fiz.\ Nauk {\bf 159}, 401 (1989)].
}

\lref\PolyakovUZ{
D.~Polyakov,
Nucl.\ Phys.\ B {\bf 449}, 159 (1995)
[arXiv:hep-th/9502124].
}

\lref\BelopolskyCY{
A.~Belopolsky,
Phys.\ Lett.\ B {\bf 403}, 47 (1997)
[arXiv:hep-th/9609220].
}

\lref\BelopolskyBG{
A.~Belopolsky,
arXiv:hep-th/9703183.
}

\lref\BelopolskyJZ{
A.~Belopolsky,
arXiv:hep-th/9706033.
}

\lref\BerkovitsRB{
N.~Berkovits,
JHEP {\bf 0109}, 016 (2001)
[arXiv:hep-th/0105050].
}

\lref\WittenZZ{ 
E.~Witten, 
hep-th/9112056. 
} 
 
\lref\wichen{ 
E.~Witten, 
arXiv:hep-th/9207094. 
}

\lref\BaranovCG{
M.~A.~Baranov, I.~V.~Frolov and A.~S.~Schwarz,
Theor.\ Math.\ Phys.\  {\bf 70}, 64 (1987)
[Teor.\ Mat.\ Fiz.\  {\bf 70}, 92 (1987)].
}

\lref\voronov{Th.~Voronov, 
Sov.~Sci.~Rev.~C Math. Phys. {\bf 9} (1992), 1-138. }

\lref\GaidukNI{
A.~V.~Gaiduk, O.~M.~Khudaverdian and A.~S.~Schwarz,
Theor.\ Math.\ Phys.\  {\bf 52}, 862 (1982)
[Teor.\ Mat.\ Fiz.\  {\bf 52}, 375 (1982)].
} 

\lref\BirminghamTY{
D.~Birmingham, M.~Blau, M.~Rakowski and G.~Thompson,
Phys.\ Rept.\  {\bf 209}, 129 (1991).
}

\lref\WittenFB{
E.~Witten,
Prog.\ Math.\  {\bf 133}, 637 (1995)
[arXiv:hep-th/9207094].
}

\lref\SiegelKU{
W.~Siegel and B.~Zwiebach,
Nucl.\ Phys.\ B {\bf 299}, 206 (1988).
}

\lref\SiegelAP{
W.~Siegel,
Phys.\ Lett.\ B {\bf 142}, 276 (1984).
}

\lref\BerkovitsGJ{
N.~Berkovits, M.~T.~Hatsuda and W.~Siegel,
Nucl.\ Phys.\ B {\bf 371}, 434 (1992)
[arXiv:hep-th/9108021].
}

\lref\AisakaGA{
Y.~Aisaka and Y.~Kazama,
JHEP {\bf 0404}, 070 (2004)
[arXiv:hep-th/0404141].
}

\lref\BerkovitsTW{
N.~Berkovits and D.~Z.~Marchioro,
arXiv:hep-th/0412198.
}

\lref\OdaBG{
I.~Oda and M.~Tonin,
arXiv:hep-th/0409052.
}

\lref\GrassiWZW{
P.~A.~Grassi, G.~Policastro and P.~van Nieuwenhuizen,
Nucl.\ Phys.\ B {\bf 676}, 43 (2004)
[arXiv:hep-th/0307056];
P.~A.~Grassi, G.~Policastro and P.~van Nieuwenhuizen,
arXiv:hep-th/0402122.
}

\lref\GuttenbergHT{
S.~Guttenberg, J.~Knapp and M.~Kreuzer,
JHEP {\bf 0406}, 030 (2004)
[arXiv:hep-th/0405007].
}

\lref\WittenMK{
E.~Witten,
Nucl.\ Phys.\ B {\bf 371}, 191 (1992).
}

\lref\AbeEP{
Y.~Abe, V.~P.~Nair and M.~I.~Park,
arXiv:hep-th/0408191.
}

\lref\Rflat{
M.~Rocek and N.~Wadhwa,
arXiv:hep-th/0408188;
C.~g.~Zhou,
arXiv:hep-th/0410047;
M.~Rocek and N.~Wadhwa,
arXiv:hep-th/0410081.
}

\lref\SiegelRX{
W.~Siegel,
Mod.\ Phys.\ Lett.\ A {\bf 5}, 2767 (1990).
}

\lref\BerkovitsZK{
N.~Berkovits,
arXiv:hep-th/0209059.
}

\lref\manin{
Yu.~I.~Manin, {\it Gauge Field Theory and Complex Geometry}, 
Springer, Berlin, 1988 (Grundlehren der Mathematischen Wissenschaften,
289).}

\lref\berleit{
J.~N.~Bernstein, D.~Leites, 
 Funct. Anal. Appl. {\bf 11}, N3 (1977) 70-71.
}

\lref\BershadskyHK{
M.~Bershadsky, S.~Zhukov and A.~Vaintrob,
Nucl.\ Phys.\ B {\bf 559}, 205 (1999)
[arXiv:hep-th/9902180].
}

\lref\BerkovitsZQ{
N.~Berkovits, M.~Bershadsky, T.~Hauer, S.~Zhukov and B.~Zwiebach,
Nucl.\ Phys.\ B {\bf 567}, 61 (2000)
[arXiv:hep-th/9907200].
}

\lref\BerkovitsYR{
N.~Berkovits and O.~Chandia,
Nucl.\ Phys.\ B {\bf 596}, 185 (2001)
[arXiv:hep-th/0009168].
}

\lref\BerkovitsZV{
N.~Berkovits,
JHEP {\bf 0204}, 037 (2002)
[arXiv:hep-th/0203248].
}

\lref\OoguriQP{
H.~Ooguri and C.~Vafa,
Adv.\ Theor.\ Math.\ Phys.\  {\bf 7}, 53 (2003)
[arXiv:hep-th/0302109].
}

\lref\deBoerDN{
J.~de Boer, P.~A.~Grassi and P.~van Nieuwenhuizen,
Phys.\ Lett.\ B {\bf 574}, 98 (2003)
[arXiv:hep-th/0302078].
}

\lref\HokerGW{
E.~D'Hoker and D.~H.~Phong,
arXiv:hep-th/0211111.
}

\lref\PeetersUB{
K.~Peeters, P.~Vanhove and A.~Westerberg,
Class.\ Quant.\ Grav.\  {\bf 19}, 2699 (2002)
[arXiv:hep-th/0112157].
}

\lref\BerkovitsZK{
N.~Berkovits,
arXiv:hep-th/0209059.
}

\lref\GrassiXR{
P.~A.~Grassi and P.~Vanhove,
arXiv:hep-th/0411167.
}

\lref\HerbstJP{
M.~Herbst, C.~I.~Lazaroiu and W.~Lerche,
arXiv:hep-th/0402110.
}

\lref\HofmanZT{
C.~M.~Hofman and W.~K.~Ma,
JHEP {\bf 0106}, 033 (2001)
[arXiv:hep-th/0102201].
}

\lref\HofmanCW{
C.~Hofman,
JHEP {\bf 0311}, 069 (2003)
[arXiv:hep-th/0204157].
}

\lref\HofmanJZ{
C.~Hofman and J.~S.~Park,
Commun.\ Math.\ Phys.\  {\bf 249}, 249 (2004)
[arXiv:hep-th/0209214].
}

\lref\HofmanRV{
C.~Hofman and J.~S.~Park,
arXiv:hep-th/0209148.
}

\lref\LazaroiuZI{
C.~I.~Lazaroiu,
arXiv:hep-th/0312286.
}

\lref\KulaxiziPA{
M.~Kulaxizi and K.~Zoubos,
arXiv:hep-th/0410122.
}

\lref\RuizRuizYF{
F.~Ruiz Ruiz and P.~van Nieuwenhuizen,
Nucl.\ Phys.\ B {\bf 486}, 443 (1997)
[arXiv:hep-th/9609074].
}

\lref\GatesNR{
S.~J.~Gates, M.~T.~Grisaru, M.~Rocek and W.~Siegel,
{\it Superspace, Or One Thousand And One Lessons In Supersymmetry,}
Front.\ Phys.\  {\bf 58}, 1 (1983)
[arXiv:hep-th/0108200].
}

\lref\GrassiPoli{
P.A. Grassi and G. Policastro, in preparation
}

\lref\ghs{
A.~S.~Galperin, P.~S.~Howe and K.~S.~Stelle,
Nucl.\ Phys.\ B {\bf 368}, 248 (1992)
[arXiv:hep-th/9201020].
}

\lref\sorokin{
D.~P.~Sorokin, V.~I.~Tkach and D.~V.~Volkov,
Mod.\ Phys.\ Lett.\ A {\bf 4}, 901 (1989);
D.~P.~Sorokin, V.~I.~Tkach, D.~V.~Volkov and A.~A.~Zheltukhin,
Phys.\ Lett.\ B {\bf 216}, 302 (1989).
}


\listtoc
\writetoc
\hrule
\vfill
\eject

\newsec{Introduction}

The interest on superstrings 
and topological strings with supermanifold as target space 
has increased during the last months. 
This was mainly due to the progresses in formulating 
the superstrings in 10 dimensions with manifest super-Poincar\'e invariance 
\refs{\BerkovitsFE,\BerkovitsUE,\GrassiUG} and in formulating 4d 
${\cal N}=4$ SYM  
in the supertwistor space \refs{\WittenNN,\BerkovitsHG,\NeitzkePF}. 
The pure spinor appoach allows 
the consistent quantization of superstrings 
for any generic background, for example with Ramond-Ramond fields 
such as
 the celebrated $AdS_{5} \times S^{5}$ (see for example \BerkovitsYR\ for 
 a worldsheet formulation), while the twistor formulation  
 allows a direct comparison between Feynman diagram computations 
 in quantum field theory and the correlation functions 
 of a topological B model. The common ground for these models 
 is the super-target space and the problems 
 of  constructing the amplitudes have the same geometrical origin: 
 the  theory of integration of superforms on supermanifolds.  
  
 The study of supermanifolds as string theory target space 
 has to be 
 traced back to the original papers by M. Green and J. Schwarz 
 \GreenZG\ where a sigma model for 
 10d superstring with manifest supersymmetry was formulated. 
 Later, A. Schwarz {\it et al.} started the analyis of topological 
 sigma models with target space as supermanifolds \SchwarzAK. 
 More recently, mirror symmetry has given a new acceleration  
 to the analysis of super target spaces \refs{\NeitzkePF,\SethiCH,
 \AganagicYH,\KumarDJ}. Indeed, the formulation 
 of mirror symmetry \refs{\HoriKT,\BookTop}
 based on Landau-Ginzburg models directly leads to supermanifolds. 
 This is due to the fact that 
 the superpotential of non-linear Landau-Ginzburg 
 theory must be modified by introducing some ghost superfields. 
 These are twisted fermion fields which, together with the 
 bosonic superfields, parametrize a supermanifold and the problem 
 of integration on such manifolds emerges again. 
  
 For a bosonic manifold ${\cal M}$, the theory of integration is 
 related to the theory of antisymmetric tensors  
 in the cotangent bundle $T^{*}{\cal M}$. 
 Indeed, the measure for a given simplex is obtained by 
 constructing the top form of the submanifold. However, 
 this cannot be naively generalized to superspaces $\hat{\cal M}$. 
 The naive space of superforms $\Omega^{*}\hat{\cal M}$ 
 can be locally factorized (if the supermanifold is split, \ie is obtained by
 the total space of a vector bundle by changing the parity of the fibres) into
 $\Omega^{*}{\cal M} \oplus  
 C^{\infty}(\Omega^{*}{\cal N})$ where ${\cal M}$ is the bosonic 
 body of the supermanifold $\hat{\cal M}$ and ${\cal N}$ is 
 the fermionic extension. Although these superforms 
 are polynomial in the anticommuting coordinated $d x^{m}$, they 
 are functions of the commuting superform $d\t^{\a}$ 
 (we choose the set of coordinates $(x^{m}, \t^{\a})$ on 
 $\hat{\cal M}$ which is locally ${\cal M} \times {\cal N}$). 
 In the purely bosonic case the 
 form degree can only be equal or less than the dimension of 
 the manifold and the top form transforms as a measure under 
 smooth, orientation preserving coordinates transformations. 
 This allows one to integrate the top forms over the oriented 
 manifold. On the other hand, superforms may have any 
 form degree and none of them transforms as a Berezinian measure. 
 Some generalizations have been proposed by Bernstein and 
 Leites \berleit\ and they introduced the concept of 
 pseudoforms as distributions on $\Omega^{*}{\cal M} \oplus 
 C^{\infty}(\Omega^{*}{\cal N})$ (in the literature they are also 
 called integral forms, see for example book \manin). 
 But how is this related to 
 superstrings and topological strings on supermanifolds? 
   
 In the pioneering paper  
 \FriedanGE\ 
 Friedan, Martinec and Shenker in the case of fermionic 
 strings (superstrings with worldsheet fermions and 2d local
 supersymmetry) pointed out that one  
 has to insert into correlation functions some 
 suitable BRST-closed operators to soak up 
 the zero modes of bosonic ghosts -- the superghosts of 
 2d supergravity. They indeed conjectured that 
 if one considers the superspace version of the worldsheet  
 (adding some fermionic coordinates) the superghosts can 
 be viewed as the differentials of the fermionic coordinates and 
 therefore the insertions, known also as Picture Changing Operators (PCO), 
 were needed in order to make a sensible integration theory for 
 superforms (the vertex operators) on the super-worldsheet. 
 Later, it was  further recognized \refs{\VerlindeSD,\KnizhnikAK,\PolyakovUZ} 
 that the gauge fixing of the gravitino of the 2d superconformal gravity was 
 directly related to the choice of those PCO insertions.  The 
 need of a gauge choice for the gravitino is 
 particularly important in the computations at higher genus 
 \refs{\HokerGW,\PeetersUB}. 
 
Fermionic strings is not the only $\sigma$-model yielding 
a supersymmetric string theory. The Green-Schwarz 
formalism indeed provides an example of a sigma model 
with supersymmetry in 
the target space and this can be quantized using pure spinor 
formalism (see \BerkovitsZK\ for a pedagogical account). 
In that framework, the construction of higher genus Riemann surface 
amplitudes  requires PCO to reabsorb  the zero modes of the 
commuting ghosts \BerkovitsZK. Recently in 
\refs{\AnguelovaPG,\GrassiXR}, it has been showed that 
rules similar to those used in \BerkovitsZK\ can be employed to provide a 
loop expansion in the point-particle limit of string theory and for 
a particle description of supergravity. In this case, the PCO 
insertions are not derived from a gauge fixing of some 
worldsheet gauge fields, and they are motivated by symmetries 
(BRST symmetry) and ghost anomaly cancellation. 
However, the bosonic ghosts $\l^{\a}$ of the pure spinor formalism 
are BRST partners of the fermionic coordinates $\t^{\a}$ of the target space 
and this suggests that $\l^{\a}$ are the differential of $\t^{\a}$ where 
the role of the de Rham differential is played by the BRST charge. The 
path integral over the zero modes of $\t^{\a}$ and $\l^{\a}$ can then be viewed 
as an integral of superforms. 
Following the 
same logic, the ghosts of the fermionic coordinates 
in supertwistor space ${\bf CP}^{3|4}$ must require 
corresponding PCO in the amplitudes. Indeed in this context the analogy 
is rather close as pointed out in footnote 12 of \WittenNN.

 The connection between PCO and supergeometry 
is studied in papers  \refs{\BelopolskyCY,\BelopolskyBG,
\BelopolskyJZ}. There the author provides a bridge between 
the formalism for integration on supermanifolds 
\refs{\SchwarzAK,\BaranovCG,\voronov,\GaidukNI} and the 
PCO. However, the application of his formalism to 
target space supermanifold in the context of superparticle and 
topological strings is lacking and it is discussed in the present work. 
We should also mention the work by Sethi \SethiCH\ where 
some brief comments on integration on supermanifolds were made.  

In order to clarify these issues, we consider a simplified model: 
we take into account a topological worldline model -- Sec. 2.1 --
(see \BirminghamTY\ for a pedagogical account) with 
the supermanifold ${\bf R}^{3|2}$ as target space (this 
is the easiest supersymmetric generalization of the model taken into
consideration  
by Witten in \WittenFB; it has N=1 supersymmetry in the target space  
parametrized by $x^{m}$ with $m=0,1,2$ and $\t^{\a}$ with
$\a=1,2$. The latter is a Majorana spinor in 3d). This model  
can be identified with the open sector of a topological A model on the
worldsheet; in this context, ${\bf R}^{3|2}$ is a Lagrangian
submanifold which defines the boundary conditions for open strings --
see Sec. 2.2.  
The particle model only deals with the zero modes 
of the worldsheet theory.  The choice of a superparticle model instead 
of a superstring is due to our interest in the zero modes of the theory and 
we therefore neglect normal ordering problems and other worldsheet 
details. A similar analysis is performed in \SiegelRX. 
The target space theory is supersymmetric Chern-Simons theory which 
is essentially given by the usual Chern-Simons theory of the a
3-manifold plus a  
mass term for non-dynamical fermions 
\lref\SchwarzYJ{
J.~H.~Schwarz,
JHEP {\bf 0411}, 078 (2004)
[arXiv:hep-th/0411077].
}
\refs{\GatesNR,\RuizRuizYF,\SchwarzYJ}. 
It is a supersymmetric model and 
the equations of motion are given by null curvatures plus vanishing fermions. 
The quantization is performed along the lines of  \refs{\GrassiUG,\GrassiXF} 
extending the work of \BerkovitsRB\ by relaxing the pure spinor
constraints. In Sec. 3, we  
discuss the cohomology of the model at any ghost number at zero and 
at non-zero momentum. The zero momentum cohomology is then used to 
characterize the tree level measure and to derive the target space 
action directly from Witten's string-field theory as in \WittenFB. 

The relation between the cohomology at different ghost numbers and 
the Batalin-Vilkovisky formalism for the off-shell theory 
was discovered in \refs{\SiegelKU,\SiegelAP} and applied 
to RNS formalism in \BerkovitsGJ. In those papers, 
the zero momentum cohomology at the highest ghost number 
provides a meaningful integration measure for tree level correlation
functions.   
In the present work (see Sec.~3.3), we compute the highest (in the 
present case is 3) ghost number 
cohomology group and we found out that there is a unique element 
which is a polynomial 
in the ghost fields and in the fermion coordinates. 
In terms of this polynomial 
we derive the tree-level measure.  
The resulting measure is BRST invariant and   
 it contains enough Dirac delta functions to reabsorb the zero modes as 
 expected. 
As discussed above, those 
delta functions should appear as insertions by 
means of PCO in order to guarantee the gauge invariance 
and the supersymmetry of the model.
For that purpose in Sec.~4, we derive the PCO from a complete 
geometrical method based on supergeometry and we show 
that in the space of pseudoforms the top form coincides with 
the measure computed with the zero momentum cohomology.  
This points out the dependence of the PCO on the  
gauge choice, and it illustrates some aspects of the formula for higher genus 
supestring correlators given in \BerkovitsPX. Indeed, we clarify the 
relation between the zero momentum cohomology and the 
path integral measure by identifying the correct top form to be integrated, 
and we develop the Cartan calculus for the supergeometry. 
We show that a gauge fixed bosonization formula 
for the commuting ghosts leads to the construction of PCO 
and the picture number operator (as will be defined later in Sec.~4.6). 
We show that the anomaly of the ghost number and of the picture number 
is saturated by the measure constructed in Sec.~5. 
 There, we finally give the prescription for multiloop computations 
with any number of external states. In Sec.~5.1, we derive the 
insertions computed in Sec.~5 by coupling the model to an extended 
topological gravity on the worldline. There are essentailly two symmetries 
that are important: the Virasoro constraint $P^{2} \sim 0$ and 
the $\kappa$-symmetry constraints $\not\!\!P d \sim 0$. We derive 
both the insertions for composite antighost field $B$ and for the picture 
changing operators. 
In app. A, we collected very few basic ingredients of Batalin-Vilkovisky 
formalism used in the text. 


\newsec{Actions and BRST Symmetry}

\subsec{Worldline action and BRST symmetry}

We consider the Liouville action (the action of the general form $\int P dq$) 
on the worldline to which gauge fields 
are added\foot{The action in (2.1) is  closely related to the 
action provided in 
\refs{\sorokin,\ghs} with the main difference that, here, it is 
supersymmetrized in the target space and the conjugate $p_{\a}$ is added 
to the theory. The latter is an independent degree of freedom. 
One can also understand the action in (2.1) as a superparticle moving 
on a supermanifolds as will be discussed at length in sec. (2.2).} 
\eqn\actA{
S= \int d\tau \Big[ P_m \Big( \dot{x}^m - l^m(\tau) \Big) -
 p_\a \Big( \dot{\t}^\a - \Lambda^\a(\tau) \Big)
\Big]\,, 
}
Here $x^m$ are the bosonic coordinates (we consider 
$D$-dimensional target space manifolds where $D=(d,1)$ and
$d=2,3,5,9$. In those particular  
cases we have the Fierz identities $\eta_{mn} \g^n_{{(\a\b}} \g^m_{\g\d)} = 0$ 
for the Dirac matrices of the corresponding spaces).
The coordinates of the corresponding superspaces are $(x^m,\t^\a)$ where 
the index $\a$ runs over $\a = 1,2$ for $d=2$ (Majorana-Weyl spinors);
$\a = 1,\dots,4$ for $d=3$ (Majorana spinors); 
$\a = 1,\dots,4$ for $d=5$ (symplectic-Majorana spinors 
\lref\six{
P.~Fayet,
Nucl.\ Phys.\ B {\bf 263}, 649 (1986)\semi
J.~Hughes, J.~Liu and J.~Polchinski,
Phys.\ Lett.\ B {\bf 180}, 370 (1986).
}
\six
) 
and finally $\a = 1,\dots,16$ for $d=9$ (Majorana-Weyl spinors). 

The action in \actA~is invariant under the following gauge
transformations with local parameters  
$\eta^{\a}$ and $\zeta^{m}$ 
\eqn\actB{\eqalign{
&\d x^m = \zeta^m + {i \over 2} (\eta \g^m \t)\,, 
~~~\d \t^\a = \eta^\a \,,
~~~ \d l^m = \dot\zeta^m + {i\over 2} \, (\eta \g^m \Lambda) -
{i\over2}(\t \g \dot\eta)\,, \cr  
&\d \Lambda^\a = \dot\eta^\a\,,
~~~ \d P_m = 0\,, 
~~~ \d p_\a =  {i\over 2} P_m (\g^m \eta)_\a\,, 
}}
and under rigid super-Poincar\'e transformations 
\eqn\actC{\eqalign{
&\d_\e x^m = a^m - {i\a\over 4}(\e \g^m \t )\,, 
~~~ \d_\e \t^\a = \e^\a\,, 
~~~ \d_\e l^m = + {i \a \over 4}  (\e \g^m \dot\t ) - { i \b \over 2}
(\e  \g^{m} \L) \,, \cr 
&\d \Lambda^\a = 0 \,,
~~~\d_\e P_m =0\,, ~~~
\d_\e p_\a = + {i \b \over 2} P_m (\g^m \e)_\a\,, 
}}
with constant parameters $a^m$ and $\e^\a$ and $\a +\b =1$. 
By redefining $l^{m} \rightarrow l^{m} + a \dot x^m$, we can set $\a =0$.
The gauge transformations remove any propagating 
degrees of freedom and, therefore, this worldline model should 
describe a topological model in target space whose physical sector 
is restricted to the zero modes \WittenFB.\foot{The main difference
with the conventional  
superparticle action $S = \int d\tau {1\over 2 e} (\dot x^m - {i \over
2} \t \g^m \dot \t)^2$  of ref. \lref\BS{
L.~Brink and J.~H.~Schwarz,
Phys.\ Lett.\ B {\bf 100}, 310 (1981).
}
 \BS\  
is that for that model the gauge symmetries are the reparametrizations
on the worldline and the $\kappa$-symmetry.} 
Notice that the 
non-linear terms in the gauge transformations \actB~such as $ {i \over
2} (\eta \g^m \t)$  in $\delta l^{m}$  
are needed in order that the gauge symmetry commutes with the super-Poincar\'e 
transformations \actC\ by assuming that the gauge parameters $\zeta^{m}$ and 
$\eta^{\a}$ are supersymmetric invariant.  

The action should also be diffeomorphism invariant; the corresponding 
transformation rules are $\d x^{m} = \xi P^{m}$,  
$\d P^{m} =0$ and $\d l^{m} = {d\over dt} (\xi
P^{m})$, but comparison with \actB\ shows that  
these transformation rules are just a special case of the gauge transformations, 
with $\zeta^{m} = P^{m} \xi$. 
Thus we do not treat diffeomorphisms  
separately. In addition, the action is also invariant 
under a fermionic symmetry (Siegel $\kappa$ symmetry) with 
$\delta_{\kappa} \t^{\a} = \g^{\a\b}_{m} P^{m} k_{\b}$ and 
$\delta_{\kappa} x^{m} = \delta_{\kappa}\t^{\a} \g^{m}_{\a\b} \dot\t^{\b}$ 
(the conjugate momenta $p_{\a}$ and $P^{m}$ transform accordingly). 
Again, this symmetry is part of the original symmetry when the 
gauge parameters are  
$\eta^{\a} = \g^{\a\b}_{m} P^{m} k_{\b}$. 
These transformations are easily compensated by changing the gauge fields 
$\Lambda^{\a}$ and $l^{m}$. It 
is interesting to note that the reparametrizations and the 
$\kappa$ symmetry are a subalgebra of the gauge transformations and 
they will play a role in the definition of amplitudes. 

To render \actA~explicitly supersymmetric, we introduce 
the composite fields $\Pi^m = \dot{x}^m + {i \over 2 } (\t \g^m \dot \t)$, 
$d_\a = p_\a + {i \over 2} P_m (\g^m \t)_\a$ and $L^m = l^m + {i \over
2} (\t \g^m \Lambda)$. All of them are supersymmetric expressions and
in terms of them the action reads  
\eqn\actD{
S= \int d\tau \Big[ P_m \Big( \Pi^m - L^m \Big) -
 d_\a \Big( \dot{\t}^\a - \Lambda^\a \Big)
\Big]\,.  
}
By replacing the gauge parameters $\xi^m$ and $\eta^\a$ by 
ghost fields $c^m$ and $\l^\a$ (which are anticommuting and 
commuting, respectively), we obtain a nilpotent BRST symmetry
\eqn\actE{
s \, x^m = c^m + {i \over 2} (\l \g^m \t)\,, 
~~~s \t^\a = \l^\a\,,
~~~ s\, L^m = \dot c^m + i\, (\l \g^m \Lambda)\,, ~~~s \Lambda^\a =
\dot\lambda^\a\,,} 
$$s P_m = 0\,, 
~~~s \, d_\a =  i  P_m (\g^m \l)_{\a}\,, ~~~
s \, c^m = - {i\over 2} (\l \g^m \l)\,, ~~~ s \l^\a = 0\,. 
$$
We notice that the BRST transformations map the fields $x^m$ and $\t^\a$ into 
fields and ghosts, the gauge fields $L^m$ and $\Lambda^\a$ into gauge
fields and  
ghosts and, finally, the composite operators $P_m$ and $d_\a$ are
mapped into themselves and  
ghosts. 

To gauge-fix the action, we introduce the anti-ghost fields $b_m$ and 
$w_\a$ and the BRST-auxiliary fields $\rho_m$ and $\rho_\a$, which 
transform under BRST symmetry as follows 
\eqn\actF{\eqalign{
s\, b_m = - \rho_m\,,~~~~ s\, \rho_m =0\,,~~~ s w_\a = - \rho_\a + i\,
(\l \g^m b_m)\,,~~~ 
s\, \rho_\a =  i\, \rho_m (\g^m \l)_\a\,.
}}
Notice that $s \, w_\a$ contains a 
nonlinear term with $i (\l \g^m b_m)$. 
We could have written the simpler transformation laws $s \hat w_{\a} =
\hat \rho_{\a}$ and  
$s \hat\rho_{\a} =0$ , but in order to have manifest supersymmetry  we
have shifted $w_{\a}$ as $\hat w_{\a} = w_{\a} + i b_{m} (\g^{m}
\t)_{\a}$, and this leads also to the nonlinear term $i \rho_{m}
(\g^{m} \l)_{\a}$ in $s \, \rho_{\a}$. 

Next, we add a gauge fermion 
\eqn\actF{\eqalign{
S' =& S + s \int d\tau \Big[ b_m \Big( L^m - \half P^m\Big) + w_\a
\Lambda^\a \Big] = \cr 
=& S + \int d\tau \left[ \rho_m \left( L^m - \half P^m \right) - b_m
\dot c^m +  
\rho_\a \Lambda^\a + w_\a \dot\lambda^\a \right] \,.
}}
Eliminating the auxiliary fields $\rho_m$ and $\rho_\a$ and the gauge
fields $L^m, \Lambda^\a$  
by their algebraic equations of motion, we arrive at the final action
and the final BRST charge 
\eqn\actG{
S'= \int d\tau \Big[ P_m  \Pi^m - \half P^2   - d_\a \dot{\t}^\a - b_m
\dot c^m + w_\a \dot\lambda^\a 
\Big]\,,}
$$
Q = \l^\a d_\a - c^m P_m + {i \over 2} \, b_m (\l \g^m \l)\,. 
$$
Fields transform as $s\, \Phi = i [Q, \Phi\}$, and this reproduces
\actE\ and \actF. 

By using the commutators $[P^m, x^n] = -i \, \eta^{m n}$, 
$\{p_\a, \t^\b \}= -i \, \d_\a^{~\b}$, $\{b_m, c^n\} = -i \, \d_m^{~n}$ and 
$[w_\a, \l^\b]=  -i \, \d_\a^{~\b}$, it is easy to show that the 
fields $d_\a$ satisfy the algebra $\{d_\a, d_\b\} = - P_m \g^m_{\a\b}$ and 
the BRST charge $Q$ is nilpotent. We take $P_{m}, 
\l^{\a}, \t^{\a}$ and $b_{m}$ hermitian, and $d_{\a}$ and $c^{m}$ 
antihermitian; then  $Q$ is antihermitian. Both the action and the 
BRST charge are invariant under supersymmetry.\foot{Note that the bosonic 
sector of the action \actG, which contains $x^m$ and the 
corresponding ghosts $c^m$ and $b^m$, resembles the conventional 
quantized bosonic point-particle whose free action is given 
$S'= \int d\tau \Big[ P_m  \dot x^m - \half P^2   - b \dot c \Big]$. 
The main difference is that the ghost $b$ and $c$ are scalar fields 
with respect to target-space Lorentz transformations.} The action 
resembles very closely the superparticle quantized with pure spinors 
\BerkovitsZK (see also \GrassiXF), apart from the presence of the 
anticommuting ghosts and the fact that the spinors $\l^{\a}$ are 
unconstrained.\foot{In \GrassiUG, the introduction of new ghosts 
classified by different grading numbers  
has led to a similar BRST charge and, in order to single out 
the physical states, one has to consider the restricted functional 
space with non-negative grading.} 
The main difference between the present model and the 
$D=(9,1)$ superstring  is that in the latter case the constraints
$\l^\a \g^m_{\a\b} \l^\b =0$  
can be solved, thereby reducing the number of independent ghost
fields. In lower dimensions , such  
as in 3,4, and 6 dimensions, there is no solution besides the trivial
one (for Majorana spinors).  
However, in the present context, the fact that there is no solution to
the constraint means that there are  
no physical degrees of freedom. Thus, we obtain  a purely topological model. 

As in topological strings, the Virasoro constraints $P^{2} = 0$ are
not present in the  
BRST charge and correspondingly also the ghost and the antighosts of the
reparametrization invariance  
are absent. Thus it is difficult to establish the correct measure for
different worldlines. But in the present context 
the relation 
\eqn\vira{
P^{2} = \{ Q, b_{m} P^{m} \} 
}
is valid. 
We therefore introduce the composite field 
$B = b_{m} P^{m}$, which plays the role of the 
antighost in the usual reparametrization invariant theory 
for the particle. Moreover, 
we can also use the reparametrizations generated by $P^{2} \sim 0$ 
to set two coordinates of $x^{m}$  to zero by choosing a light-cone gauge. 
However, this choice would break the manifest supersymmetry. To avoid 
this we must invoke another symmetry to remove consistently two 
fermionic dof's. This can be done by another gauge symmetry  generated  
by $\not\!P d \sim 0$ which forms  
a subalgebra together with $P^{2}$. 
Its BRST transformation 
$\{Q, \not\!P d\} = \l^{\a} P^{2}$  does not vanish\foot{ 
In pure spinor string theory, there is no negative ghost number operator; due 
to the gauge invariance generated by the pure spinor constraints,
$w_{\a}$ appear  
only in gauge invariant combinations $w \g^{mn}\l$ and
$w_{\a}\l^{\a}$, which have  
ghost number zero. In that case there is an operator $b^{\a}$ such
that $\{Q, b^{\a}\} = P^{2} \l^{\a}$,  
which has been used in \BerkovitsPX\ to define the amplitudes on genus
$g$ Riemann surfaces.} and therefore $\not\!\!P d \sim 0$  cannot 
be put on the same ground as $P^{2}$. Nevertheless, we can 
modify $\not\!P d \sim 0$ by adding $\l^{\a} b_{m} P^{m}$ to 
render it BRST close and, in addition, BRST exact. 
We postpone the discussion on the implications at the end of 
Sec. 5.  

Although it is very convenient to work with 
supersymmetric quantities, we have to point out 
that there is a unitary transformation which suitably simplifies the
BRST charge 
\eqn\actH{
Q' = e^{-\half \l^\a \g^m_{\a\b} \t^\b b_m} Q  e^{\half \l^\a
\g^m_{\a\b} \t^\b b_m} 
= \l^\a p_\a - c^m P_m\,.
}
The new charge is still nilpotent, but it is not supersymmetric.
The form of the BRST symmetry in \actH\ may be convenient for 
other purposes, but we use $Q$ given in \actG.\foot{We acknowledge 
W. Siegel for a discussion on the relation 
between the present formalism and the 
``Big Picture'' formalism \BerkovitsGJ.}

\subsec{Worldsheet actions and BRST symmetry}

In the previous section, we analysed a simple model of a superparticle. In the 
present section we point out that this model can  
indeed be interpreted as the reduction of a topological string on a supermanifold. 
Here, we simply sketch the construction 
of topological strings on supermanifold. We do not pursue this analysis 
in the present paper and it will be interesting to see all these 
ideas been completely developed for A/B models, such the construction of 
$A_{\infty}$ algebra 
\HerbstJP, open/closed interactions 
\refs{
\HofmanZT,
\HofmanRV,\HofmanCW,
\LazaroiuZI,\HofmanJZ,\KulaxiziPA} and the analysis 
of boundary conditions. We postpone these issues to a future 
publication \GrassiPoli.

The model consists of maps $\Phi: \Sigma \rightarrow X^{m|n}$, from a
two-dimensional surface  
$\Sigma$ to a Riemannian supermanifold $X^{m|n}$ of bosonic 
dimension $m$ and fermionic dimension $n$, equipped with a 
supermetric $g$. If we pick local coordinates $z,\bar z$ on $\Sigma$ and 
$x^{I}, \theta^{\a}$ (where $I=1,\dots,m$ and $\a=1,\dots, n$) on  $X^{m|n}$, 
then $\Phi$ can be described locally via functions $(x^{I}(z,\bar z), 
\theta^{\a}(z,\bar z))$. Let us introduce the fields 
$(\psi^{I}_{+}, q^{\a}_{+})$, 
a section of $K^{1/2} \otimes \Phi^{*}(T X^{m|n})$ (where $T X^{m|n}$ is 
the complexified tangent bundle of $X^{m|n}$ and $K^{1/2}$ is the 
spin bundle over $\Sigma$). In the same way, we introduce 
$(\psi^{I}_{-}, q^{\a}_{-})$, a section of $K^{1/2} \otimes \Phi^{*}(T
X^{m|n})$.  
The action is 
\eqn\acA{
S = 2 t \int d^{2}z \Big( 
{1\over 2} g_{IJ} \p_{z} x^{I} \p_{\bar z} x^{J} + 
g_{I \a} \p_{z} x^{I} \p_{\bar z} \t^{\a} + 
{1\over 2} g_{\a \b} \p_{z} \t^{\a} \p_{\bar z} \t^{\b}  + 
$$
$$
+ {i\over 2} g_{IJ} \psi^{I}_{-} D_{z} \psi_{-}^{J} 
+ {i\over 2} g_{I\a} \psi^{I}_{-} D_{z} q_{-}^{\a}  
+ {i\over 2} g_{\a\b} q^{\a}_{-} D_{z} q_{-}^{\b}  
$$
$$
+ {i\over 2} g_{IJ} \psi^{I}_{+} D_{\bar z} \psi_{+}^{J} 
+ {i\over 2} g_{I\a} \psi^{I}_{+} D_{\bar z} q_{+}^{\a}  
+ {i\over 2} g_{\a\b} q^{\a}_{+} D_{\bar z} q_{+}^{\b}  
\Big) 
$$
$$+ ~{curvature ~terms}\,.
}
The supermetric $g = (g_{(IJ)}, g_{(I\a)}, g_{[\a\b]})$ has been 
decomposed along the bosonic and fermionic components. The 
super-Riemann tensor $R$ is also decomposed into several 
components. The structure of the first line recalls the 
structure of superstrings in Ramond-Ramond background 
\refs{\BerkovitsYR,\BershadskyHK,\BerkovitsZQ,
\BerkovitsZV,\OoguriQP,\deBoerDN} where the components 
of the metric $g$ along the fermionic components $g_{I\a}$ are 
proportional to the gravitinos and $g_{\a\b}$ are proportional the 
the inverse of the RR field strenghts. To see this, it convenient 
to decompose the coordinates into $(x^{i}, \t^{a}, x^{\bar i}, \t^{\bar a})$  
and introduce new fields $p_{a}$ and $p_{\bar b}$. Then, 
the first line of \acA\ can be rewritten as follows
\eqn\acB{
\hat g_{i\bar j} \p_{z} x^{i} \p_{\bar z} x^{\bar j} + 
p_{a} \p_{\bar z} \t^{a} + p_{\bar b} \p_{z} \t^{\bar b} + 
(p_{a} - g_{a \bar j} \p_{\bar z} x^{\bar j}) 
g^{a\bar b} 
(p_{\bar b} - g_{i \bar b} \p x^{i})\,.
}
The last term has the same structure of pure spinor 
string theory on $AdS_{5}\times S^{5}$ constructed in 
\BerkovitsYR. Notice that there are additional terms coming 
from the curved manifold that contributes to the curvature terms in 
\acA. This analogy with $AdS$ space might be useful to gain 
some new results in amplitude computations on a curved manifolds 
for superstrings. As in the bosonic case, we expect the model \acB\ to be 
conformally invariant if the target space is a super-Calabi-Yau manifold. 
This condition is apparently weaker than super-Ricci flatness, as has been 
recently remarked in \Rflat. 

We specialize now the previous  
formulae to the case of a complex super-3 fold ${\bf C}^{3|4}$. 
The fermionic directions
are parametrized by a symplectic Majorana spinor. 
By a suitable choice of the background (choice of D-branes and fluxes) 
it should be possible to reduce  
further the number of target space fermions to a single Majorana 
spinor of 3d and the superspace becomes ${\bf R}^{3|2}$. Most likely
the mechanism for the reduction would involve choosing a Lagrangian
submanifold of $X$.  
Then the action for the zero modes should reduce to \actG . 
We do not derive this fact here but will work with this assumption in
mind. 
Since we can give consistent rules for the amplitudes (sec. 5), we can {\it a
posteriori}   
justify this assumption, but clearly a deeper investigation is needed 
\GrassiPoli. 

As is well-known, when the target space is a K\"ahler manifold the
action \actA\ has $N=2$ worldsheet supersymmetry. The fermions can be twisted
using the fermion number  
current $U(1)$, and for the topological A model, one ends up with 
the sections $\chi^{i},\chi^{\bar i}, \l^{a}$ and $\l^{\bar a}$ and 
with the 1-forms $\psi^{i}_{\bar z}$, $\psi^{\bar i}_{z}$, $w^{\bar a}_{z}$ 
and $w^{a}_{\bar z}$. As part of the twisted N=2 supersymmetry 
we can derive the BRST transformations 
\eqn\BRSTA{
\delta x^{i} = i (\chi^{i} + \l^{a} \g_{ab}^{i} \t^{b}) \,, ~~~~~
\delta \chi^{i} = -  \l^{a} \g^{i}_{ab} \l^{b}\,, 
}
$$
\delta \t^{a} = i \l^{a}\,, ~~~~~ \delta \l^{a} =0\,,
$$
$$
\delta \psi^{\bar i}_{z} = - \p_{z} x^{\bar i} + \l^{\bar a} \g^{\bar
i}_{\bar a \bar b}  
w^{\bar b}_{z} -i  
\Big(  \chi^{\bar j} \G^{\bar i}_{\bar j \bar m} \psi^{\bar m}_{z} +  
\l^{\bar a} \G^{\bar i}_{\bar a \bar m} \psi^{\bar m}_{z} +
\chi^{\bar j} \G^{\bar i}_{\bar j \bar b} w^{\bar b}_{z} +
\l^{\bar a} \G^{\bar i}_{\bar a \bar b} w^{\bar b}_{z}
\Big) \,,
$$
$$
\delta w^{\bar a}_{z} = - \p_{z} \t^{\bar a}  - i  
\Big(  \chi^{\bar j} \G^{\bar a}_{\bar j \bar m} \psi^{\bar m}_{z} +  
\l^{\bar a} \G^{\bar a}_{\bar a \bar m} \psi^{\bar m}_{z} +
\chi^{\bar j} \G^{\bar a}_{\bar j \bar b} w^{\bar b}_{z} +
\l^{\bar a} \G^{\bar a}_{\bar a \bar b} w^{\bar b}_{z}
\Big) \,,
$$
They are nilpotent only on the equations of motion of 
the ghost fields. Using these BRST transformations, one 
can rewrite the action in the following form
\eqn\acC{
S = 2 t \int_{\Sigma} d^{2}z \{Q, V\} + 
\int_{\Sigma} x^{*}(K)
}
and the last term is the pull-back of the K\"ahler 2-form 
\eqn\KK{
K = g_{i\bar j} dx^{i} dx^{\bar j} + g_{a\bar b} d\t^{a} d\t^{\bar b}
}
and the off-diagonal terms are removed by a $\t$-dependent 
diffeomorphism on the manifold. 

Even though we are mostly concerned in this paper with the point-particle
limit, it is still useful to have the complete theory at hand. Indeed, as we
will see in section 5, the rules for calculating the amplitudes cannot
be completely justified without appealing to the conformal field
theory formalism. 

Recently, the relation between WZW model based on a supergroup 
and twistor spaces have been explored \AbeEP. 
It is worth pointing out 
that the pure spinor formulation can be seen to emerge from a 
WZW model based on supergroup \refs{\GrassiWZW,\GuttenbergHT
} and the computation rules can be established on the basis of 
\WittenMK. 


\newsec{The $D=(2,1), N=1$ Chern-Simons model}

Now we go back to the worldline model. 
In this section we compute the vertex operators, both integrated and 
unintegrated, and the zero momentum cohomology needed to construct 
the tree level measure. 

\subsec{Vertex operators}

To show that the formalism constructed in the previous section
describes a super-Chern-Simons theory, we compute the BRST  
cohomology: $\{Q, U^{(1)}\} =0$  with  $\d U^{(1)} = [Q, \Omega]$. 
The fields of the field theory, namely the gauge field $a_m$ and the
gaugino $\psi^\a$, are identified with the BRST cohomology classes at
ghost number $1$.  

Applying the BRST charge to the most general superfield with ghost
number one expressed in terms of a maximal set of commuting coordinates  
\eqn\brsAA{
U^{(1)} = \l^\a A_\a(x,\t) - c^m A_m(x,\t) \,,
} 
we have the following field equations
\eqn\brsA{
\{ Q , U^{(1)} \}  = - {i\over 2} (\l \g^m \l) A_m - c^m c^n \p_n A_m 
+ c^m \l^\b D_\b A_m - c^m \l^\b \p_m A_\b + \l^\a \l^\b 
D_\a A_\b = 0\,,}
where $D_\a = \p_\a + { i\over 2} (\g^m \t)_\a \p_m$, 
which imply 
\eqn\brsB{\eqalign{
F_{\a\b} \equiv D_{(\a} A_{\b)} - {i \over 2} \g^m_{\a\b} A_m = 0\,,~~~~~ 
F_{\a m} \equiv D_\a A_m - \p_m A_\a = 0\,, ~~~~~ 
F_{mn} = 0\,,
}}
where $F_{[mn]} = \half(\p_m A_n- \p_n A_m)$. 
The last equation is a consequence of the first 
two equations by exploiting the Bianchi identities 
$[\nabla_m, \{ \nabla_\a, \nabla_\b\} ] + \{ \nabla_\a, [\nabla_\b,
\nabla_m] \} - 
 \{ \nabla_\b, [\nabla_m, \nabla_\a] \}=0$ where $\nabla_\a = D_\a +
 A_\a$ and   
$\nabla_m = \p_m + A_m$. 
Equations \brsB~are invariant under $ \d U^{(1)} = [Q, \Omega^{(0)}]$, or, in
 terms of the  
components $A_\a, A_m$ of the connections,  $\d A_\a = D_\a 
\Omega^{(0)}$ and $\d A_m = \p_m \Omega^{(0)}$. 

From now on we consider the case of 3-dimensional target space: $x=
x^m$ with $m=0,1,2$. Then,  
$\t^\a \t^\b = - \half \e^{\a\b} (\t^\a \t_\a)$ where $\t^{\a} =
\t^{\b} \e_{\b\a}$.  
Decomposing the superfields $A_\a$ and $A_m$ in the following way, 
\eqn\brsC{\eqalign{
A_m = a_m + \t^\b \hat\xi_{m\b} + ( \t^\a \t_\a ) \xi_m\,, ~~~~~~ 
A_\a = \chi_\a + \t^\b \hat\chi_{\a\b} + \t^\b \t_{\b} \psi_\a\,.
}}
and eliminating the auxiliary fields $\hat\xi_{m\b},  \xi_m,  \chi_\a, 
\hat\chi_{\a\b}$ by using the field equations and the algebraic gauge
transformations,  
one finds the equations of motion $\p^m a_n - \p_n a_m =0$ and
$\psi^\a =0$. These coincide with the equation of motion for
super-Chern-Simons theory. After removing  
the auxiliary fields, the supersymmetry is realized by the usual
transformation laws  
$\delta_\e a_m = (\e \g^m \psi)$ and $\delta_\e \psi^\a = \half F_{mn}
(\g^{mn}\e)^\a$.  

Another important element is the integrated vertex operator with 
integrand $V$. 
This satisfies the equation $\{Q, V^{(0)}\} = \dot U^{(1)}$, and 
it is defined up to the gauge transformations $\delta V^{(0)} = \dot \Omega^{(0)}$, 
where $\Omega^{(0)}$ is the superfield of the gauge transformations. 
For on-shell fields $\dot \t^{\a}  = 0, \dot P_{m} =0, \dots$ 
the vertex $V$ has the generic form
\eqn\VV{
V^{(0)} = \dot x^{m} A_{m} + d_{\a} W^{\a} + 
w_{\a} \l^{\b} F^{\a}_{~\b} + b_{m} c^{n} F^{m}_{~n} + 
w_{\a} c^{m} F^{\a}_{~m} + b_{m} \l^{\a} F^{m}_{~\a} \,.
}
However, imposing $\{Q, V^{(0)}\} = \dot U^{(1)}$ and using the equations of
motion \brsB (which implies that all the curvatures vanish), 
the vertex reduces  to 
\eqn\VVV{
V = \dot x^{m} A_{m} \,.
}
This coincides exactly with the usual vertex operator in the 
case of bosonic Chern-Simons in D=(2,1). One can finally 
define the Wilson loop by setting 
\eqn\WL{
W(\g) = {\rm tr} \Big( P e^{\int_{\g} d\tau \dot x^{m} A_{m}} \Big)\,,
}
where $\g$ is a curve in the target space. $P$ denotes the path 
ordering and the trace is then needed for non-abelian gauge-invariance. 
Notice that dealing with non-abelian gauge 
group, the superderivatives in \brsA\ and \brsB\ 
should be replaced by covariant 
superderivatives. In the following section the complete non-abelian 
action will be discussed. Furthermore, $\{ Q, W(\g) \} =0$, and it
cannot be written  
as a BRST exact quantities. As is well-known no local gauge invariant 
observable can be constructed for Chern-Simons theory. Because $A_{m}$ is 
a superfield, $W(\g)$ has an expansion in terms of zero modes $\t^{\a}$. 

In order to study the target space field theory, 
it is useful to compute also the cohomologies at 
ghost number $0,2,3$ and higher. It is easy to see that the equation 
$\{ Q , U^{(0)} \}  =0$ for a ghost number zero superfield
$U^{(0)}$ implies  
that it is a constant. Usually the cohomology at ghost number 2 
doubles the cohomology at ghost number 1 
if the momentum does not vanish (see for example \BerkovitsGJ). 
The most general superfield at ghost number 2 is given by 
$U^{(2)}= 
\l^\a \l^\b A^*_{\a\b} + \l^\a c^m A^*_{\a m} + c^m c^n A^*_{m n}$,
and clearly  
the number of antifields exceeds the number of corresponding superfields 
$A_\a$ and $A_m$. Therefore, we expect that the equations of motion
will show that  
some of the antifields of $A^*_{\a\b}, A^*_{\a m}$ and $A^*_{m n}$ are 
redundant and can be expressed in terms of the others. 

From BRST invariance $\{ Q, U^{(2)} \} =0$, we
obtain the following equations 
\eqn\brsC{\eqalign{
& D_{(\a} A^*_{\b\g)} + {i \over 2} \g^m_{(\a\b} A^{*}_{\g)m} = 0\,,~~~~~~
D_{(\a} A^*_{\b)m} - {i \over 2} \g^n_{\a\b} A^{*}_{[mn]} + 
\p_m A^*_{(\a\b)}= 0\,, \cr
& D_{\a} A^*_{[m n]} - \p_{[m} A^*_{|\a| n])}= 0\,, ~~~~~~
\p_{[m} A^*_{n r]} = 0\,. 
}}
We can decompose the superfields $A^*_{(\a\b)}, A^*_{\a m}$ and
$A^*_{m n}$ into  
irreducible representation of the super-Poincar\'e group: $A^*_{\a\b}
= \g^m_{\a\b} B^*_m$,  
$A^*_{\a m} = \g_{m\a\b} B^{*\b} +  B^*_{\a m}$, where $B^*_{\a m}$ is 
$\g$-traceless and $A^*_{m n} = \e_{mnr} B^{* r}$. From the equations
of motion,  
one obtains that $B^*_{\a m}$ and $B^{* r}$ are algebraically related to 
one vector $B^*_m$ and one spinor superfield $B^{*\b}$. Those two superfields 
are the antifields for $A_m$ and $A_\a$. The gauge transformations which 
leave the equations of motion of the antifields $A^{*}$ invariant
should be the equations of motion of the superfield  
$A$.
This is indeed the case since $\delta U^{(2)} = Q \Omega^{(1)}$
with $\Omega^{(1)} = \l^{\a} C_{\a} - c^{m} C_{m}$ yields  
\eqn\brsD{\eqalign{
&\d A^*_{(\a\b)} = D_{(\a} C_{\b)} - {i\over 2} \g^m_{\a\b} C_m\,, ~~~~
\d  A^*_{\a m} = D_\a C_m - \p_m C_\a\,,\cr
&\d A^*_{[m n]} = \half ( \p_m C_n - \p_n C_m) \,, 
}}
where $C_m$ and $C_\a$ are two arbitrary superfields. 

For what concerns the ghost-number three BRST cohomology, we  
point out that the only solution is a constant scalar field for 
non-zero momentum cohomology. One expects that there is no
cohomology at ghost number beyond three. The easiest way to show this
is to first determine the zero-momentum cohomology  
because the latter contains the case $k^m \neq 0$ as a special case,
while it is easier to compute. So we now turn to the zero-momentum
cohomology.  

\subsec{Zero momentum cohomology and tree level measure}

We denote here with $\Psi$ the string field without any restriction 
on the ghost number. 
Direct evaluation of $\{Q, \Psi\} =0$ with $Q = \l^{\a} \p_{\a} + 
{i\over 2} b_{m} \l \g^{m}\l$ 
shows that the most general solution for the 
cohomology at zero momentum is given by\foot{
For example, at ghost number one the equations to solve are 
$\p_{\a} A_{m} = 0$ and $\p_{(\a} A_{\b)} - {i \over 2} \g^{m}_{\a\b} A_{m} 
= 0$, which yield $U^{(1)} = \l^{\a} A_{\a} - c^{m}A_{m} = \l^{\a} \Big( 
{i\over 2} \g^{m} _{\a\b} a_{m} \t^{\b} \Big)- c^{m} a_{m}$ 
with constant $a_{m}$. }
\eqn\zmc{
\eqalign{
&\Psi = c \, U^{0} + a_m U^{(1) m} + a^{*,m}  U^{(2)}_{m} + c^* U^{(3)}\,, \cr
& U^{(0)} =1\,, \cr
& U^{(1),m} = {i\over 2} \l^\a \g^m_{\a\b} \t^\b - c^m \,, \cr
& U^{(2)}_{m} = \e_{mnr} \left(  {1\over 4} \l \g^n \t  \,  \l \g^r \t  -i
\l \g^n \t  \, c^r + c^n \, c^r \right)\,, \cr 
& U^{(3)} = \e_{mnr} \left(  {3\over 2} \l \g^m \t  \,  \l \g^n \t \, c^r
- { 3 i}  
\l \g^m \t  \, c^n \, c^r + c^m \, c^n \, c^r \right)\,. 
}}

To show that $Q$ annihilates $U^{(3)}$ we used the identity $\t^{\a} \t^{\b} = 
- {1\over 2} \e^{\a\b} \t^{2}$, and $[\g^{m}, \g^{n}] = 2 i \e^{mnr} \g_{r}$. 
The only remaining degree of freedom is the gauge vector $a_m$, which 
cannot be removed at zero momentum by a gauge transformation 
$\delta \Psi = \{Q, \Omega\}$. On the 
contrary, the gaugino can be removed by a gauge transformation 
at zero momentum.   
The constant fields $c$ and $c^*$ are the ghost and its antifield, and they  
are gauge invariant. 

The path integral measure can be decomposed into two factors 
$d\mu = d\mu_{0} d\widetilde{\mu}$, where $d\mu_{0}$ is the 
measure on zero modes and $d\widetilde{\mu}$ is the measure on 
non-zero modes. The latter is chosen as usual in the conventional way, the 
free measure weighted with classical action evalutated on non-zero modes. (In 
conformal field theory this part of the measure is obtained by
performing all the  
possible OPE's among the insertion of vertex operators). Since the 
action for the zero modes vanishes we should choose a different path to 
define it. 

The path integral measure for zero modes is defined in the following 
way 
\eqn\defPIM{
\int d\mu_{0} U^{(3)} = 1
}
where $U^{(3)}$ is an element of the cohomology $H^{3}(Q)$. This 
follows the usual requirement that the integral measure is the 
Poincar\'e dual to the top form here represented by the highest 
non-vanishing element of the BRST cohomology. The main property 
of this measure is $\int d\mu_{0} Q\Lambda =0$ for any $\Lambda^{2}$. 
The total ghost number of $d\mu_{0}$ is $-3$. In the 
following sections, we will show that this number is associated 
to the anomalies of $U(1)$ currents. 

According to this definition, it is easy to see that 
\eqn\eva{
\langle U^{(0)} U^{(3)} \rangle = \int d\mu_{0} U^{(0)} U^{(3)} = constant\,,  ~~~~~~
\langle U^{(1)} U^{(2)} \rangle = \int d\mu_{0} U^{(1)} U^{(2)} =constant\,. 
}
The cohomology $U^{3}$ in \zmc\ is composed by three independent 
monomials  and we have to establish the integration rules for each of them. 
Thus we need 
\eqn\corr{\eqalign{
&\langle c^m \l^\a \l^\b \rangle = \e^{mpq} \g^{\a\a'}_p \g^{\b\b'}_q
\p_{\a'} \p_{\b'} \,,\cr 
& \langle c^m c^n \l^\a \rangle = u\, \e^{mnq} \g^{\a\a'}_q \p_{\a'} \,, \cr
& \langle c^m c^n c^r \rangle = w\, \e^{mnr} \,.
}}
where $u$ and $w$ are arbitrary constants. 
The path integral measure associated with the correlators \corr\ is given by 
\eqn\measu{\eqalign{
&\langle F(x, \t, \l,c) \rangle = \int d\mu_{0} F(x,\t, \l,c) \,, \cr
&d\mu_{0} = \Big( \e^m_{~nr} c^n c^r \g^{\a\b}_m  \p_{\l^\a}  \p_{\l^\b} 
- u (\t \not\! c \, \p_{\l}) + w \, \t^2 \Big) \d^2(\l)\, d^3x \,d^2
\t\, d^2 \l\,  d^3 c\, 
}}
where $\p_{\l}$ are derivatives which act on the delta function $\d^2(\l)$. 
The presence of delta functions is due to bosonic ghost fields 
and we have to reabsorb their zero modes to have a non-vanishing and 
non-divergent contribution. In sec. 4, it is illustrated how this
measure is related to the usual bosonic measure of Chern-Simons theory
$c^{3}$.  
To 
fix the free parameter $u$ and $w$, we adopt the following strategy: we check 
which term provides a supersymmetric amplitude and we check the gauge
invariance.  

\item{$\triangleright$} Supersymmetry. 

Notice that the last term of the functional measure gives the bosonic 
Chern-Simons action which is not supersymmetric. In fact, as is clear
from \measu~ the last term does not provide a supersymmetric measure.  
The function $F(x, \t, \l,c)$ is a polynomial combination of
superfields and transforms as  
$\delta_\e F(x, \t, \l,c) = \e^\a D_\a F(x, \t, \l,c)$ under
supersymmetry. Integrating by parts, one has  
\eqn\measuB{\eqalign{
\langle \d_\e F(x, \t, \l,c) \rangle =  \e^\a 
\int d^3x \,d^2 \t\, d^2 \l\,  d^3 c\,   \left(  D_\a  \mu_{0}
\right)\,  F(x,\t, \l,c) =  
0\,.
}} where $d\mu_{0} = \mu_{0}  d^3x \,d^2 \t\, d^2 \l\,  d^3 c$. 
Since $\left(  D_\a  \mu_{0}\right) \neq 0$, we set 
$u=w=0$. 

\item{$\triangleright$} Gauge invariance.

Another essential property for the measure is gauge invariance. 
Namely, the integration of a BRST trivial vertex should vanish:
$\langle \{Q,\Omega\}  \rangle =  
\int d\mu_{0} \{Q, \Omega\} = 0$. In fact, inserting 
the vertex 
$Q( \l^\a \l^\b \Omega_{(\a\b)} + \l^\a c^m \Omega_{\a m} + c^m c^n
\Omega_{[mn]} )$ and  
selecting only  terms of the form $\l^\a \l^\b c^m$ because $u=w=0$, 
\eqn\gaugeINV{
\Big\langle - \l^\a \l^\b c^n \p_n \Omega_{(\a\b)} + 
\l^\a c^m \l^\b D_\b \Omega_{\a m} + i \, \l^\a \l^\b c^n \g^{m\a\b}
\Omega_{[mn]} \Big\rangle =}  
$$=\int d^{3}x d^{2}\t \g^{n\a\b} \Big(  \p_n \Omega_{(\a\b)} +
D_{(\b} \Omega_{\a) n}  +  
\g^{m}_{\a\b}  \Omega_{[mn]} 
\Big) =0\,.
$$
The first term vanishes because it is a total derivative, 
the second because it is a total spinorial derivative and the 
last one because ${\rm tr}(\g^m \g^n) = 2 \eta^{mn}$ and $\Omega_{[mn]}$ 
is antisymmetric.\foot{The measure \measu\ can be rewritten in the
following form 
\eqn\meA{
\langle F \rangle = 
\int  d^3x \,d^2 \t\, d^2 \l\,  d^3 c \int d^{2}w \, 
\e_{\a\d} (\not\!c w)^{\a} (\not\!c w)^{\d} 
e^{ w_{\a} \l^{\a}} F 
}
where we introduced the integral representation for the delta function. 
 The last line in the above equation can again be written as 
 an exponent by introducing a new fermionic zero mode $p_{\a}$
\eqn\meB{
\langle F \rangle = 
\int  d^3x d^{3} c\, d^2 \t\,  d^{2}p \, d^2 \l\, d^{2}w \, 
e^{ (p_{\a} \g^{\a\b}_{m} \,w_{\b} \, c^{m} + w_{\a} \l^{\a} ) } F \,.
} A similar analysis has been pursued in \GrassiWZW\ where we first point out 
the necessity of delta functions in the path integral of $\l^{\a}$.} 
This is not unexpected since, as will be shown in sec. 4, the delta 
functions apper in the PCO which are BRST closed. 
  
\subsec{The action}

It is convenient to use a Witten-like string field theory 
to derive the action. Moreover, this derivation leads to 
a BV action which 
contains the classical action plus all the fields and 
antifields needed to implement the symmetries of the model. 
For that we need a supersymmetric BV 
measure $\omega_{BV}$  (see app.~ A for further details) 
which reads 
\eqn\coC{\eqalign{
\omega_{BV} = \langle \Psi, \Psi \rangle = 
\int d^3x \int d^2\t \, \g^{m\a\b} \left( \d A^*_{(\a\b)} \d A_m +  
\d A^*_{\a m} \d A_\b + \d C^*_{\a\b m} \, \d C 
\right)\,.
}}

The BV action for the fields and antifields which appear in the most general 
string field
\eqn\coB{\eqalign{
\Psi &= C + \l^\a A_\a + c^m A_m + \l^\a \l^\b A^*_{\a\b} + \l^\a c^m
A^*_{\a m} +  
c^m c^n A^*_{[m n]}  \cr 
& +\l^\a \l^\b \l^\g C^*_{\a\b\g} + \l^\a \l^\b c^m C^*_{(\a\b) m}  + 
\l^\a c^m c^n C^*_{\a [m n]} + c^m c^n c^r C^*_{[m n r]} \,.
}}
should couple the fields $A_{\a}$ and $A_{m}$ to the corresponding 
antifields $A^{*}_{\a\b}, A^{*}_{\a m}$ and $A^{*}_{mn}$, and the 
ghost $C$ to the corresponding antifields $C^{*}_{\a\b\g}, C^{*}_{\a\b
m}, C^{*}_{\a mn}$ and  
$C^{*}_{mnr}$. (From the zero momentum cohomology, we know that there
is only one non-vanishing  
state corresponding to a scalar superfield $C^* = \g^{m \a\b} C^*_{(\a\b) m}$.)

The string action $S$, which satisfies the master equation, is obtained from 
\eqn\CS{S=  \half \langle \Psi , Q \, \Psi \rangle + 
{1\over 3} \langle  \Psi , \Psi \Psi \rangle \,,}
where $Q$ is the BRST charge given \actG~and the generic state is 
described in \coB. Again the product in the interaction term 
is given by $\langle \Psi_1 \Psi_2 \Psi_3 \rangle = 
\int  \mu(\l,c,\t)  {\rm tr} \left( \Psi_1\Psi_2 \Psi_3 \right)$. This 
vanishes in the abelian case. In the non-abelian case, the string
field $\Psi$ carries an  
index in the adjoint representation of the gauge group and 
${\rm tr} \left( \Psi_1\Psi_2 \Psi_3 \right)= f_{abc}
\Psi^{a}_1\Psi^{b}_2 \Psi^{c}_3$.  
The inner product by given by 
\eqn\innerprod{
\langle \Psi_1, \Psi_2 \rangle = 
\int \mu(x,\l,c,\t)  \left( \Psi_1\Psi_2 \right)} 
where the product $\left( \Psi_1\Psi_2 \right)$ is 
the usual superspace product of superfields. Notice that in the
non-abelian case, the  
superfield product should include the trace over the internal gauge group: 
${\rm tr} \left( \Psi_1\Psi_2 \right)$. 
The fields $C^*_{\a\b\g}, \dots, C^*_{[m n r]}$ are the 
antighost fields for the ghost $C$. 

In string field theory the master equation is defined  by $(S,S) = 
\int \mu {\delta S \over \delta \Psi}{ \delta S \over \delta \Psi}
$. Substitution of  
$S$ in \CS\ yields $\int \mu (Q \Psi + \Psi^{2} )(Q \Psi +
\Psi^{2})$. This vanishes  
because $\int \mu  Q\Psi Q\Psi = \int Q( \Psi Q\Psi) = 0$, while $\int \mu 
(Q\Psi) \Psi^{2} = {1\over 3}\int Q (\Psi^{3})$, and finally  $\Psi^{4} = 0$ 
because $\Psi$ is anticommuting. Subsitution of the measure yields 
the more familiar result:
\eqn\bvmA{
\{ S, S\}_{BV} = 
\int d^3x\int d^2\t \, \g^{m \a\b}
\left( 
{\p_l S \over \p A^*_{(\a\b)}}  {\p_r S \over \p A_{m} } +  
{\p_l S \over \p A^*_{\a m}}  {\p_r S \over \p A_{\b} }  + 
{\p_l S \over \p C^*_{(\a\b) m}}  {\p_r S \over \p C}
\right) = 0 \,. 
}
 
Inserting the expression of $\Psi$ and of the BRST charge, from \CS\ using the 
measure $\mu$ given in \measuB\ one 
obtains the supersymmetric invariant action
\eqn\CSA{\eqalign{
&S = \int d^3x \int d^2\t \g^{m\a\b} {\rm tr} X_{m\a\b} \,, \cr
&X_{M N R} = \half A_{[M} D_N A_{R\}} + {1\over 3}  A_{[M} A_N A_{R\}} + 
A^*_{[MN} D_{R\}} C + C^*_{[MNR\}} C^2  
}}
where the index $M$ refers to both the vector index $m$ and the
spinorial index $\a$,  
and $D_{M} = \{\p_{m}, D_{\a} \}$. 
To check that \CSA~reproduces exactly the Chern-Simons action, we can 
compute the kinetic terms, neglecting both the contributions coming 
from the non-abelian terms and from the antifields. The form 
of the action has also been described in \SiegelRX.


Assuming the decomposition 
\eqn\supe{
\eqalign{
A_m = a_m + \t^\b \hat\xi_{m\b} + ( \t^\a \e_{\a\b} \t^\b ) \xi_m\,,
~~~~ A_\a = \chi_\a + \t^\b \hat\chi_{\a\b} + \t^2 \psi_\a\,. 
}}
we have for the first term in \CSA\foot
{We use the conventions: ${\rm tr}(\g^m \g^n \g^p) = -2 \e^{mnp}$,
$\e^{mnp} \e_{mqr} =  
\d^n_q \d^p_r - \d^n_r \d^p_q$, $\e^{mnp} \e_{mnq} = 2! \, \d^p_q$ and
$\e^{mnp} \e_{mnp} = 3! $.  
As always, one has ${\rm tr}(\g^m) =0$, $(\g^m \g^n \g^r)_{\a\b} =
\g^m_{\a\b} \eta^{nr} - \g^n_{\a\b} \eta^{mr} +  
\g^r_{\a\b} \eta^{mn} + \e^{mnr} \e_{\a\b}$. The tensor $\e_{\a\b}$ is
normalized as $\e_{12} = 1$. Notice  
that $\t^\a \t^\b = - \half \e^{\a\b} \t^2$.  
}
\eqn\pezzocll{\eqalign{
\Big\langle c^m \l^\a \l^\b &\left( -\half \g^n_{\a\b} A_m A_n + A_m
D_\b A_\a - A_\a D_\b A_m + A_\a \p_m A_\b \right) 
\Big\rangle = \cr
& \int d^{3}x \,\e^{mpq} \Big[ - 2 \e_{pq}^{~~n} ( \xi_m a_n + \xi_n
a_m ) - (\hat\xi_m \, \g_p \, \g^n \, \g_q \, \hat\xi_n)  
- 4\,  \xi_m {\rm tr} ( \g_q \hat\chi \g_p) \cr
&- 2 a_m {\rm tr}( \g_q \p_p \hat\chi) + a_m {\rm tr}(\p_s \hat\chi \,
\g_p \g^s \g_q)  
- 2 (\hat\xi_m \g_q \p_p \chi) \cr
&+ 4 (\hat\xi_m \g_p \g_q \psi) + (\hat\xi_m \g_p\g^s\g_q \p_s \chi) - 
4 (\psi \g_q \g_p \p_m\chi) \cr 
& - {\rm tr}(\g_q \hat\chi) \p_m {\rm tr}(\g_p \hat\chi) + {\rm
tr}(\hat\chi \g_p \p_m \hat\chi \g_q) 
\Big]\,,
}}

The action is invariant under the gauge transformations
\eqn\gt{\eqalign{ 
&\d \Psi = \l^\a \d A_\a + c^m \d A_m = Q \left( \Omega + \t^\a
\omega_\a + \t^2 \eta \right) \,, \cr 
&\d \chi_\a = \omega_\a \,,~~~~~~\d \hat\chi_{\a\b} = \half
\g^m_{\a\b} \p_m \Omega + 2 \e_{\a\b} \eta\,, \cr 
&\d \psi_\a = - {1\over 4} \g^m_{\a\b} \e^{\b\g} \p_m \omega_\g \,,
~~~~~~ \d a_m = \p_m \Omega \,, \cr 
&\d \hat\xi_{m\b} = \p_m \omega_\b\,, ~~~~~~ \d \xi_m = \p_m \eta\,.
}}
Moreover, since the gauge transformations of $\chi_\a$ and of the
scalar  
part of $\hat\chi_{\a\b} = \half \g^m_{\a\b} \chi_m + \e_{\a\b} \chi$,
namely $\chi$, are  
pure shifts one can remove these fields from the action by setting
them to zero.  
Hence, one gets
\eqn\action{\eqalign{
S = 
\int d^{3}x \, \e^{mpq} & \Big[ - 2 \e_{pq}^{~~n} ( \xi_m a_n + \xi_n
a_m ) - (\hat\xi_m \, \g_p \, \g^n \, \g_q \, \hat\xi_n)  
- 4\,  \xi_m {\rm tr} ( \g_q \hat\chi \g_p) \cr
&- 2 a_m {\rm tr}( \g_q \p_p \hat\chi) + a_m {\rm tr}(\p_s \hat\chi \,
\g_p \g^s \g_q) + 4 (\hat\xi_m \g_p \g_q \psi)  \cr 
&- {\rm tr}(\g_q \hat\chi) \p_m {\rm tr}(\g_p \hat\chi) + {\rm
tr}(\hat\chi \g_p \p_m \hat\chi \g_q) \cr 
}}
Eliminating the auxiliary fields $\xi_m$ and $\hat\xi_{m\b}$, one finds the 
super-Chern-Simons action 
\eqn\ac{
S = 
-6 \int d^{3}x  \Big( \e^{mpq} a_m \p_p a_q + 3 \psi^2 \Big)\,.}
This proves the fact that the measure chosen in the present 
derivation is supersymmetric and gauge invariant. 



\newsec{Supergeometry, Picture Changing Operators and 
BRST Cohomology}

The discussion in the previous section concerning the construction of
a measure for the zero-modes integrals can be better understood when
considered in the broader context of the theory of differential forms and
integration on supermanifolds. 
This framework permits a clear discussion about PCO 
and supergeometry for topological theories 
with supermanifolds as target space. This subject has a vast 
literature and we can direct the reader to the book \manin\ for 
a summary of results. In order to discuss our application, 
we discuss the integration on supermanifolds by means of 
pseudoforms and the Baranov-Schwarz transformations, then 
we construct the superforms for our topological model, we 
derive the Cartan calculus on the superforms, and finally we construct
the PCO.  
This technique could be applied to the RNS superstrings, where the
superforms are  
represented by distributions of the superghosts associated with 
the worldsheet local supersymmetry, 
and part of the analysis was performed in \BelopolskyBG.

\subsec{Integration on Supermanifolds: Pseudoforms and Densities}

One of the main problem in the superform differential 
analysis is the construction of a consistent theory of integration. 
This has been deeply analized by Schwarz, Voronov and collaborators 
\lref\sv{
Papers by Schwarz, Voronov et al. 
} (for a more extensive discussion see \voronov\ ), 
where the construction of singular superforms and their 
relation with the usual integration on manifold 
have been worked out. Here we consider some aspects of their
formulation. We will not dwell on the issue of the precise
definition of a supermanifold, which can be found in several
places (a good general reference is \manin).  

As soon as the problem was posed, it became clear that the 
theory cannot be developed very far without departing from the analogy
with the usual theory for bosonic manifolds. In particular, the main
problem is that the straightforward generalization of a differential
form does not have the right properties for integration on
submanifolds. Several different solutions have been proposed, using
various generalizations of differential forms. 

The easiest ones to describe and work with are the
{\it pseudodifferential forms} (or simply pseudoforms) of
Bernstein-Leites \berleit\ . Given a supermanifold $X^{M|N}$ (the exponent
denotes as usual the bosonic/fermionic dimension), a pseudoform is a
distribution on the cotangent space $T^*X$. In practice then it is a
generalized function $\omega(z, dz)$ of the coordinates $(z^A) = (x^m,
\theta^\a)$ and their differentials $(dx^{m}, d\t^{\a})$. 
It can be integrated on $T^*X$
if it decreases sufficiently fast at infinity in the bosonic
directions along the fibres. On the other hand the integration along a
submanifold cannot always be defined, because in general it would
involve a product of distributions (or said differently: unlike a
function, a distribution cannot always be restricted to a subspace). 
Nevertheless, since they are easier to manipulate, 
we will work mainly with the
complex of pseudoforms, denoted by $\Omega^*(X) = 
\sum_{m,n} \Omega^{m|n}(X)$ (where the range of the summation 
will be clarified later), after establishing
their relation with two other types of objects, called {\it densities}
\GaidukNI\ . 

The densities are constructed to be integrated over submanifolds, and so they
come with a (bi)grading corresponding to the superdimension of the
submanifold. 
We define the $m|n$-{\it densities} $\CD^{m|n}(X)$  to be functions
$A(v_1,\ldots v_m, w_1, 
\ldots w_n)$ of $m$ even and $n$ odd vector fields. At each point of
$X$, the arguments of $A$ span an $m|n$ dimensional subspace of the
tangent space $TX$. Writing a matrix $R$ having the vectors $(v_i, w_j)$
as columns, this subspace is just the image of $R$, seen as a linear map
$$R:{\bf R}^{m|n} \rightarrow TX\,.$$ 
So, without any ambiguity, we write the density $A$ as $A(R)$. 
Under a linear change of basis $L \in
GL(m|n)$, a density should transform as $A(R L) = A(R) \, {\rm Ber} \,
L$, where ${\rm Ber}$ is the berezinian of a supermatrix defined as 
follows 
$$Ber(L) = {Det(L_{1} -L_{3} L_{4}^{-1} L_{2}) \over Det L_{4}}$$ where 
$L_{i}$ are the four blocks of a supermatrix. 
Recalling the properties of the 
Berezinian one can see that densities are homogeneous of degree -1 in
the odd vectors, and therefore will necessarily be singular if the odd
vectors are not linearly independent (\ie if $R$ does not have maximal
rank in the fermionic subspace). 
Given a submanifold $Y \subset X$ of
dimension $m|n$, in terms of a parametrization $z^A = f^A (\zeta^K)$ 
where the index $K$ runs over bosonic and fermonic indices, it is
possible to define the integral over $Y$ of a density:
$$ \int_Y d\zeta^K A\Big(f(\zeta), {\partial f^A \over \partial \zeta^K}\Big) \,.$$ 
This is well-defined thanks to the transformation properties of $A$. 
Notice that a density of degree equal to the dimension of the manifold
is the same as a volume form on the manifold. In the notations we are
using, this is defined by $A(R) = {\rm Ber} \, R$. Of course, other
volume forms can be obtained by multiplication with a function $f(z)$,
and the integral over the whole space will be non-vanishing only when
$f = g(x) \t^1 \ldots \t^N$. Then it is natural to single this one out
as a preferred volume form $vol_X$. 

There are also D-{\it densities} $\CDD^{m|n}$, defined as functions of
$M-m|N-n$ 
cotangent vectors, or of an operator $$S: TX \rightarrow
{\bf R}^{M-m|N-n}.$$ Such an operator has an $m|n$-dimensional kernel
which has to be thought of as the tangent space of a submanifold. If we
require the transformation property  $B(TS) = B(S) \, {\rm Ber} \, T$,
for $T \in GL(M-m|N-n)$, then the following integral is well defined: 
$$
\int_X B \Big(z, {\partial \Phi^K \over \partial z^A} \Big) \, \prod_K
\delta(\Phi^K(z)) dz^A \,.
$$
Thus D-densities can be integrated over a submanifold defined in terms
of equations $\{ \Phi^K(z) = 0, K= 1, \ldots m|n \}$. 

Clearly the concepts of densities and D-densities are equivalent, and
there is a canonical isomorphism $\CD^{m|n} \simeq \CDD^{M-m|N-n}$
given by $B(z,S) = A(z, S^\dagger)$. However, we can do more if $X$ as
a metric: we can define an Hodge duality. First of all, note that if
$R$, $S$ have maximal rank, 
we can assume, after a change of basis, that they are isometries
on the image and kernel respectively. Then is sufficient
to define a density on such isometries and extend it to a general
operator using the covariance properties. 
Given $R$, there is an orthogonal decomposition $TX = V \oplus W$, 
with $V={\rm Im} \, R$. Then $W \simeq {\bf R}^{M-m|N-n}$, and one
defines $S$ as 
the orthogonal projection over $W$. Then setting $B(S) = A(R)$, we
define a 1-1 correspondence $\CD^{m|n} \simeq \CDD^{m|n}$. Combining this
with the natural isomorphism given above, we see that 
$$\CD^{m|n} \simeq \CD^{M-m|N-n} \,.
$$
This is what we expect from Hodge duality. We need however one more
step: the classical formula which defines Hodge dual in the bosonic
case is  
$$ \a \wedge * \b = (\a,\b) \, vol_X \, .$$
We would like this to hold with the preferred choice of the volume
form that has non-vanishing integral. This uniquely defines the *
operation in our case. 

The relation between densities and pseudoforms on a supermanifold $X$ 
is expressed by the Baranov-Schwarz transformation. It is given as a 
collection of maps 
\eqn\barschwar{
\l^{m|n} : \Omega^* \to \CD^{m|n}\,, }
$$
 \omega(z^A,dz^A) \mapsto  [\l^{m|n}\omega](z^{A}, dz^{A}) =
 A(z,R) \equiv \int_{{\bf R}^{m|n}} D(dt^F) \, \omega(z^A, dt^F R_F^A) \,.
$$
To avoid confusion, we notice that here $D(dt)$ denotes Berezin
integration over the variable $dt$.  
A pseudoform does not have a grading in principle. But if its image
under the transformation is zero except for one $\l^{m|n}$ then
it can be said to be homogeneous of degree $m|n$. We can then
formulate Hodge duality 
for homogeneous pseudoforms. 
To see how this works explicitly, we consider the simplest possible
case, namely $X = {\bf R}^{1|1}$, 
with coordinate $(x, \theta)$. A vector on $TX \simeq X$ will be written as
$v=(v^1,v^{\bar 1})$. Take the pseudoform $\omega = dx$. Its 
transform in $\CD^{1|0}$, a function of one even vector, is given by:
$$A(v) = \Big[\lambda^{1|0} \omega\Big] (v) = \int D(dt) \, dt \, v^1 = v^1 \,.
$$
We find now the dual form in $\CD^{0|1}$, a function of an odd vector 
$w = (w^{1}, \tilde w^{\bar 1})$ with $w^{1}$ anticommuting and 
$\tilde w^{\bar 1}$ commuting:
\eqn\dual{
\tilde A (w) = \tilde A \left( \matrix{w^1 \cr w^{\bar 1}}  
\right) = {1 \over w^{\bar 1}} \tilde A \left( \matrix{w^1/w^{\bar 1} \cr
1}  \right) = {1 \over w^{\bar 1}} B({w^1 \over
w^{\bar 1}}, 1) = {1 \over w^{\bar 1}} A \left( \matrix{1 \cr
w^1/w^{\bar 1}} \right) = {1\over w^{\bar 1}} \,.
}
In the first equality we used the linearity of the $\tilde A$. Then we use 
the properties of the BS transformations. 
It is easy to see that this is the Baranov-Schwarz transform of
$\delta(d\theta)$ (in fact $[\l^{1|0}\omega](v) = \int D(dt)
\delta(\tilde w^{\bar 1} dt) =  
\int D( \tilde w^{\bar 1}dt) \delta(\tilde w^{\bar 1} dt) /{\tilde
w}^{\bar 1} = {1\over {\tilde w}^{\bar 1}}$.)  

If we start instead with $\omega=d\theta$, 
$$A(v) = \lambda^{1|0}\omega (v) = \int D(dt) \, dt \, v^{\bar 1} =
v^{\bar 1} \,.
$$
The computation of the dual form is the same as in \dual\ , except at
the last step when $A$ now picks up the fermionic component of the
argument, so  
$$\tilde A(w) = {w^1 \over (w^{\bar 1})^2} \,. $$
This is the BS-transform of $\omega = dx \delta'(d\theta)$, in fact 
\eqn\dualb{\lambda^{0|1} \omega (w) = \int D(dt) w^1 \, dt \,
\delta'(w^{\bar 1} dt) = {w^1 \over (w^{\bar 1})^2} \,.}
In conclusion, we find the action of * in $\Bbb{R}^{1|1}$ to be 
\eqn\star{\eqalign{
*(dx) & = \t \, \delta(d\t) \cr
*(d\t) & = dx \, \t \, \delta'(d\t) \,.}}
Applying again the BS transform, it is easy to show that $*^{2} = 1$. 

We come now to the case of interest to us, $X={\bf R}^{3|2}$. 
We will compute the dual of the forms that appear in the measure
 \measu\ . We use now notations compatible with those of the previous
 section, which amounts to making the identifications $dx^m \to c^m$,
 $d\t^\a \to \l^\a$. 
Let us start with the last term, $\omega = \t^2
\delta^2(\l)$ and we consider a 
vector of $T^{3|2}X$ denoted by $(w^{m}, w^{\a})$. 
Its BS transform in $\CD^{0|2}$ is $A(w^\a) = {\rm
det}^{-1} (w^\a_\b)$. 
The dual form $A \in \CD^{3|0}$ is 
\eqn\dualc{
\tilde A(v^{m}) = \tilde A \left( \matrix{v^m_n \cr v^m_\a} \right) =
{\rm det}(v^m_n) \tilde A \left( \matrix{ 1 \cr v^m_\a (v^{-1})_m^n}
\right) = {\rm det}(v^m_n) A(w_\a) = {\rm det}(v^m_n) \,,
}
where $w_\a$ are odd vectors with $w^\a_\b =\delta^\a_\b$, and 
$w^\a_m = v^\a_n(v^{-1})^n_m$. 
Then $A(w_\a) = 1$ from which the last equality
follows. Clearly $\tilde A$ is the BS transform of the pseudoform
$\epsilon_{mnp} c^m c^n c^p$. 
The other terms are less easy. We will explicitly compute only the
first one, $\omega=\epsilon_{mnr} c^m c^n \gamma^{r \a\b} {\partial
\over \partial \l^\a } {\partial \over \partial \l^\b } \delta^2
(\l)$. Its transform in $\CD^{0|2}$ is 
$$ A(w_\a) = {\det}^{-1}(w_\a^\b) \epsilon_{mnr} w_\rho^m
(w^{-1})^\rho_\a \gamma^{r\a\b} (w^{-1})_\b^\s w_\s^n \,.$$
In computing the dual, we have the same relation as before between
$v_m$ and $w_\a$, and using the last expression in \dualc\ we get 
\eqn\duald{
\tilde A(v_m) = {\det}(v_i^j) \epsilon_{mnp} (v^{-1})^m_r (v^{-1})^n_s
v^r_\a \gamma^{p\a\b} v^s_\b = v^r_\a \gamma^{p\a\b} v^s_\b
\epsilon_{rst} v_p^t \,.}
Noting that the expression is linear in the bosonic part and bilinear
in the fermionic part of $v_m$, we can conclude that this is the
BS transform of a pseudoform which is schematically $c \l \l$. A
careful computation shows 
that $ * \omega = \epsilon_{mnr} c^m (\theta \gamma^n \l)
(\theta \gamma^r \l)$. 

We can now understand the origin of the measure \measu\ : its terms are
precisely the duals of the terms generated from $\omega_0 = \epsilon_{mnp} \hat
c^m \hat c^n \hat c^p$, by the unitary transformation \actH, which
acts on the fields as follows:  
\eqn\newfield{\eqalign{
\hat p_\a & = p_\a +  {i \over 2} b_m (\gamma^m \lambda)_\a \,, \cr
\hat c_m & = c_m + {i \over 2} (\l \g_m \t) \,, \cr
\hat w_\a & = w_\a - {i \over 2} b_m (\g^m \t)_\a \,.
}}
In the next section we study in detail the complex of pseudoforms and
its cohomology, and we will show that $\omega_0$ is the unique element
of the cohomology of $Q'$ in degree $3|0$. 

\subsec{Superforms}

We have learnt that we can use the BS transform of superforms to define a 
meaningful integration on supermanifold. So, in the following we will 
consider only the pseudoforms and we describe them. Afterwards we discuss the 
BRST complexes of these forms and we show how the PCO operators 
play a role in the present framework. 

In the case of the superspace  ${\bf R}^{3|2}$ , the complex of
superforms contains the following spaces: 
\eqn\rsA{\eqalign{
& \Omega^{r|0} \neq 0\,, ~~~~~ {\rm for}~~ r \geq 0\,, \cr
& \Omega^{r|1} \neq 0\,, ~~~~~ {\rm for}~~ r \in {\bf Z}\,, \cr
& \Omega^{r|2} \neq 0\,, ~~~~~ {\rm for}~~ r \leq 3\,, 
}}
and the spaces are empty for other values of $r|s$. Here $s$ counts the number 
of delta functions in the pseudoform. Again, instead of
using the superform  
notation $dx^{m}$ and $d\t^{\a}$, we replace them with 
the ghosts $c^{m}$ and $\l^{\a}$ which have the same statistics. 

In order to 
write covariant expressions involving $\delta(\l)$, we 
have two possibilities: $\delta^{2}(\l)$ and $\delta(v_{\a} \l^{\a})$ 
where $v_{\a}$ is a spinor. In general, $v_{\a}$ is not constant 
and
 to define an  expression with a single delta function, 
we have to choose a direction in the spinorial space 
(represented by the real spinor 
$v_{\a}$), but 
we should cover the complete space. So, we introduce two 
of such spinors $v^{(i)}_{\b}$ where $i=1,2$ labels the 
elements of a basis of the vector space and we define the 
two delta functions as $\delta(v^{(i)}_{\b} \l^{\b})$. In addition, 
the basis is orthonormal 
$v^{(i)}_{\g} \e^{\g\d} v^{(i)}_{\d} =  \e^{ij}$. This reduces 
the number of indipendent real components of $v^{(i)}_{\b}$ to 
three and 
they belong to $SL(2, {\bf R})$. We want to reduce further the 
number of independent components to two, therefore we 
factorize the compact subgroup $U(1)$. We assume therefore 
that the coordinates $v^{(i)}_{\b}$ belong to the 
space $SL(2, {\bf R)} /U(1)$ which is not compact. They essentially  
form a set of harmonic coordinates and 
therefore in order to render the computation covariant 
at the end we integrated over the coset.\foot{Be aware that the 
integration over non-compact coordinates 
belonging a coset can be done by the Haar measure, but 
some behavior at infinity has to be assumed for the space 
of functions of the coordinates $v^{(i)}_{\b}$.
} One interesting 
choice is $v^{(i)}_{\b}(P) = \g_{m}^{i \g} \e_{\g j} P^{m} /\sqrt{P^{2}}$ 
which has $\det ( v^{(i)}_{\b} )= 1$ and where we factorize the 
$SO(2)$ rotation in the $x-y$ plane. The vectors $v^{(i)}$ 
can be viewed as gauge fixing parameters needed to fix the zero mode 
gauge (the action is clearly invariant under any transformations of the 
zero modes) and, by integrating over them, the Lorentz covariance is 
reestablished \BerkovitsPX.  
 
Explicitly, the $\Omega^{{r|0}}$ forms have the structure 
\eqn\rzero{
\Omega^{r|0} = f(x,\t) (c)^{l} (\l)^{r - l}\,, ~~~~ 0 \leq l \leq 3\,,
~~~~ r \geq 0\,.   
}
for they have  no delta functions (the vertex operator $U^{(1|0)} = \l^{\a}
A_{\a} - c^{m}A_{m}$  
discussed in Sec. 3.2 belongs to this space); for one delta function 
vertex operators 
\eqn\runo{
\Omega^{r|1} = g(x,\t) (c)^{l} (v_{\bot} \cdot\l)^{k} \delta^{(k + l
-r)}(v \cdot \l)\,, 
~~~~ 0 \leq l \leq 3\,, ~~~~ k \geq 0\,, ~~~~ r \geq 0\,.  
}
for the 1-picture forms, where $v_{\bot}$ is the orthogonal 
direction to $v$. Here, we used
\eqn\rzeroA{
\delta^{(n)}(\l^{\b}) = 
\p_{\l^{\a_{1}}} \dots \p_{\l^{\a_{n}}} \delta(\l^{\b})\,, ~~~~ n
>0\,, ~~~~~{\rm and}~~~~~ 
\delta^{(0)}(\l^{\b}) = \delta(\l^{\b}) \,.
}
The indices of the commuting ghosts $\l^{\a}$ in front of  the 
Dirac delta functions and of 
argument of the latter are different. For instance, 
$
\l^{\a} \delta^{(0)}(\l^{\b}) = \epsilon^{\a\b} \l^{\a}
\delta^{(0)}(\l_{\a})$ in such a 
way that it does not vanish. We also use the rule 
$\l^{\a} \delta^{(n)}(\l^{\a}) = (-) \delta^{(n-1)}(\l^{\a})$ where 
$\a$ is not summed. 

Finally, the two-delta function forms
are given by 
\eqn\rdue{
\Omega^{r|2} = h(x,\t) (c)^{l} \Big[\delta^{2}(\l)\Big]^{(l -r)}\,,
~~~~ 0 \leq l \leq 3\,, ~~~~  r \leq 3\,.  
}
The superscript over the delta function means $l-r$ derivative 
of the delta whit the conventions 
$\p_{\l^{\a}} \delta^{2}(\l) = 
\e_{\b\g} \Big(\p_{\l^{\a}} \delta(\l^{\b}) \Big)\delta(\l^{\g})$. 
The functions $f,g$ and $h$ are superfields. In the specific 
example of superforms that we are considering, we take into 
account only the sector with $\l^{\a}, \t^{\a}, x^{m}, c^{m}$ as 
variable and they span the manifold 
${\cal M} = {\bf R}^{3|2}\oplus T^{*}{\bf R}^{3|2}$. 

Notice that forms on ${\cal M}$ have  
only positive pictures. Later we will enlarge the space and have
negative pictures as well.


Let us discuss the de Rham
cohomology in the space of superforms. We consider first the simplest
example of $R^{1|1}$,  
with coordinates $(x,\theta)$, and differential $d = c {\p \over \p x}
+ \l {\p \over \p\t}$. The space is the product of the ordinary real
line and the 1-dimensional odd  
superspace $R^{0|1}$, and the differential splits in a sum of
operators acting independently on the two factors, so we can compute
the cohomology separately in each factor. For $R^{1|0}$, this is the
usual de Rham cohomology, but we have to pay some attention to the
behavior at infinity. If we work in the complex of forms which decay
at infinity, then the cohomolgy is generated by \foot{We denote by $\{ \dots
\}$ the span of the elements inside the brackets.} $\{1, \, c
\delta(x) \}$. We will have to consider also the zero-momentum
cohomology, defined in the complex of $x$-independent forms. Then the
generators are $\{ 1, \, c \}$. 
In the odd direction, the forms can have zero or one delta function. 
Without any delta function, 
the most general form is $\alpha = (a + b \theta)\, \l^n
\in \Omega^{n|0}$. It is closed iff $b=0$, whilst the term
proportional to $a$ is exact since    
$\l^n = d(\t \, \l^{n-1})$, so the only cohomology class is $1 \in
H^{0|0}$. At picture 1, the most general form is $\alpha = \sum_{n
\geq 0} (a_n + b_n \theta ) \delta^{(n)}(\l) \in \oplus_{n \geq 0}
\Omega^{-n|1}$.    
It is closed iff $b_n = 0, \, n \geq 1$; moreover $\delta^{(n)}(\l) =
d(\theta \delta^{(n+1)}(\l)$. So the only  
generator is $\t \delta(\l) \in H^{0|1}$. The total cohomology is
obtained by tensoring the ones  
for the two factors, and so it is  
\eqn\derhama{H^*(R^{1|1}) = \, \{1, \, c, \, \t
\delta(\l), \, c \,\t \delta(\l) \}.}  
It is interesting to see that $H^{*|0}$ and $H^{*|1}$ are isomorphic
spaces. We will see that the isomorphism can be given explicitly by
the picture changing operators. 
It is now easy to generalise from ${\bf R}^{1|1}$ to ${\bf
R}^{3|2}$. We can again use a K\"unneth-type argument to see that the
cohomology at the various pictures is given by:
\eqn\derham{\eqalign{
H^{*|0}({\bf R}^{3|2},d) = & \{ 1, c^m, c^m c^n, \epsilon_{mnp} c^m c^n
c^p \} \,, \cr
H^{*|1}({\bf R}^{3|2},d) = & \{ 1, c^m, c^m c^n, \epsilon_{mnp} c^m c^n
c^p \} \otimes \{ v^{(i)} \cdot \theta \, \delta(v^{(i)} \cdot \lambda)
\} \,, \cr 
H^{*|2}({\bf R}^{3|2},d) = & \{ 1, c^m, c^m c^n, \epsilon_{mnp} c^m c^n
c^p \} \otimes \{\theta^2 \delta^2(\lambda)\} \,. 
}}
 
We are ultimately interested in the cohomology of the complex \rsA\ with the
differential given by the BRST operator as in \actG\ . We
can observe that at zero momentum the two differentials coincide, and
moreover, as previously noticed, there is a unitary transformation \actH\ that
brings the BRST operator into the de Rham one. But it can be
convenient to work out directly the BRST cohomology in order to have
manifestly supersymmetric expressions. 
For instance, the most general vertex operator at ghost number $+1$ 
has the form
\eqn\vertnew{
U^{1|2} = c^{m} A_{m}
\delta^{2}(\l) + c^{m} c^{n} A_{nm}^{~~\a} \p_{\l^{\a}} \delta^{2}(\l) + 
c^{m}c^{n}c^{p} A_{pnm}^{~~~\a\b} \p_{\l^{\a}} \p_{\l^{\b}} \delta^{2}(\l)\,.
}
One can check that for $A_{nm}^{~~~\a} = A_{pnm}^{~~~~\a} =0$
and $F_{mn} = \p_{[m}A_{n]} =0$, this vertex is in the BRST 
cohomology. However, it will be easy to see that vertex can 
be obtained by acting on it with suitable differential operators which 
map the complesses of pseudoforms transversally. Notice that both 
$Q$ and $d$ map 
$\Omega^{r|s} \rightarrow \Omega^{r+1|s}$, and neither of them changes 
the number of delta functions. 

In fact, as will become clear later, to define the integration measure
for multiloops, we have to enlarge the complex \rsA\ to include forms
that do not have a direct geometrical interpretation as form living in
the target space. Once we make the identification of the differentials 
with the ghosts of the conformal field theory, we have to take into
account the presence of antighosts. We are then led to consider 
the full space ${\cal M}\oplus \hat{\cal M}$ where 
$\hat {\cal M}$ is the variety described by the conjugate momenta 
$P_{m}$ and $p_{\a}$ and the vector fields 
$w_{\a}$ and $b_{m}$ belonging to $T_{*}({\bf R}^{3|2} \oplus T^{*}
{\bf R}^{3|2})$.  
The variables $(x^{m},\t^{\a}, c^{m},\l^{\a})$ parametrize
${\bf R}^{3|2} \oplus T^{*} {\bf R}^{3|2}$, while $(P_{m},
d_{\a},b_{m}, w_{\a})$  
parametrize the space  $T_{*}({\bf R}^{3|2} \oplus T^{*} {\bf
R}^{3|2})$. Vectors in the latter space   
can be identified with the differential operators $(\p_{m}, D_{\a},
\iota_{m},\iota_{\a})$ 
with 
\eqn\multA{
w_{\a} \equiv \p_{\l^{\a}} = {\p \over \p ({d\t^{\a}})} =
\iota_{v_{(\a)}}\,,~~~~~~ 
b_{m} \equiv \p_{c^{m}} = {\p \over \p ({dx^{m}})} = \iota_{v_{(m)}}\,,~~~~~~
}
where 
$v_{(\a)} = v_{(\a)}^{\b} D_{\b} + v_{(\a)}^{m} \p_{m}$, 
with $v_{(\a)}^{\b} = \delta_{\a}^{\b}$ and $v_{m}^{\b} = 0$, while  
$v_{(m)} = v_{(m)}^{\b} D_{\b} + v_{(m)}^{n} \p_{n}$, with 
$v_{(m)}^{\b} = 0$ and $v_{m}^{n} = \delta_{m}^{n}$. 
Furthermore, we have 
\eqn\multB{
d_{\a} + b_{m} (\g^{m} \l)_{\a} = 
\Big[Q, w_{\a}\Big] \equiv \Big[d, \iota_{v_{(\a)}}  \Big] =  
{\cal L}_{v_{(\a)}}\,,}
$$
{\cal L}_{v_{(\a)}} = D_{\a} + (\l \g^{m})_{\a} \iota_{v_{(m)}}\,, 
$$
$$
P_{m} = 
\Big\{Q, b_{m}\Big\} \equiv 
\Big\{d, \iota_{v_{(m)}}  \Big\} =  
{\cal L}_{v_{(m)}}\,, 
$$
$$
{\cal L}_{v_{(m)}} = \p_{m}\,.
$$
The differential operator $\iota_{v_{(\a)}}$ (in the following 
we denote this operator as $\iota_{\a}$)
is commuting in contrast to the usual interior multiplication 
provided by $\iota_{v_{(m)}}$ (it will be denoted by 
$\iota_{m}$) which is anticommuting. Therefore, 
as for the superforms $\l^{\a}$, we can consider the 
superforms of negative picture of type $\delta^{2}(w) = 
\delta^{2}(\iota_{\a})$ and in general we can 
consider forms like these:  
\eqn\negPP{
\Omega^{r|-2} = m(x,\t) (c)^{l} (b)^{r} \Big[\delta^{2}(w)\Big]^{l-r-s}
}
For instance, we see that the space $\Omega^{0|0}$ of forms with
vanishing picture will contain also an element  $\delta^{2}(\l)\delta^{2}(w)$. 

\subsec{Cartan Calculus and PCO}

Given an odd vector field 
$\tilde v = 
v^{\a}(x,\t) D_{\a} + v^{m}(x,\t) \p_{m}
\equiv :v^{\a}(x,\t) d_{\a} + v^{m}(x,\t) P_{m}: 
$\foot{In the operator-formalism expressions 
in $\tilde v$ and $\iota_{\tilde v}$, 
we used the normal ordering in the 
case that the component of $v^{\a}$ and $v^{m}$ depend upon 
$x$ and $\t$. 
}
where $v^{\a}$ and $v^{m}$ are, respectively, 
commuting and anticommuting functions of $x, \t, \l$ and $c$, 
we define 
\eqn\pcoA{
\iota_{\tilde v} 
= v^{\a} \p_{\l^{\a}} + (v^{m} + v^{\a} \g^{m}_{\a\b} \t^{\b} ) \p_{c^{m}}
\equiv v^{\a} w_{\a} +  (v^{m} + v^{\a} \g^{m}_{\a\b} \t^{\b} ) b_{m} 
\,,}
$$
{\cal L}_{\tilde v} = [d, \iota_{\tilde v}] \equiv 
\Big[Q,  v^{\a} w_{\a} +  (v^{m} + v^{\a} \g^{m}_{\a\b} \t^{\b} ) b_{m} \Big]
\,,
$$
where $\iota_{\tilde v}$ is a commuting 
differential operators action on the space 
of pseudoforms $\Psi^{r|s}$. For an even vector $v$, 
the usual rules apply. The first expression in \pcoA\ is evaluated on functions 
which depend only on $\l^{\a}$ and $c^{m}$ and they are 
independent of their derivatives. Notice that the operator 
$\iota_{\tilde v}$ reduce the form degree and equivalently 
it reduces the the ghost number. However, both the operations
$Q$ and $\iota_{\tilde v}$ do not change the number of the delta functions. 
Since $Q$ is anticommuting and $\iota_{\tilde v}$ is commuting, 
we used the commutator to define the Lie derivative. 

The Cartan algebra is respected:
\eqn\CA{
Q^{2} = 0\,, ~~~~ \{Q, {\cal L}_{\tilde v} \} = 0\,, ~~~~
\{{\cal L}_{\tilde v}, \iota_{\tilde u}\} = \iota_{\{ \tilde v, \tilde
u\}}\,, ~~~ 
[\iota_{v}, \iota_{\tilde u}] = 0
}
where $\tilde u$ is another odd vector. 

Indeed ${\cal L}_{\tilde v}$ 
is an anticommuting differential operator and has the 
explicit expression
\eqn\pcoB{
{\cal L}_{\tilde v} =  - : v^{\a} d_{\a} - v^{m} P_{m} + 
\Big[\l^{\a} D_{\a} v^{\b} + c^{m} \p_{m} v^{\b}
 \Big] w_{\b}  :
 }
 $$+
:\Big[\l^{\a} D_{\a} v^{n} + c^{m} \p_{m} v^{n} + 
\l^\a D_{\a} v^{\b} \g^{m}_{\b\g} \t^{\g} + 
2(\l \g^{m} v)  \Big] b_{n}:  \,.
$$
Since the differential $\iota_{\tilde v}$ is 
commuting it can be used to define 
a formal Dirac delta function, its derivatives
and the Heaviside function 
\eqn\pcoC{
\delta(\iota_{\tilde v}) = \int_{-\infty}^{\infty} dt e^{i t
\iota_{\tilde v}}\,, ~~~~~ 
\delta^{(n)}(\iota_{\tilde v}) 
= \int_{-\infty}^{\infty} dt \, t^{n} e^{i t \iota_{\tilde v}} \,, ~~~~~
\Theta(\iota_{\tilde v}) = 
\int_{-\infty}^{\infty} dt {1\over t} e^{i t \iota_{\tilde v}} \,,
}
which are well-defined operations on the space 
of pseudoforms $\Omega^{r|s}$.  The above 
expressions can be rewritten in terms of 
$w_{\a}$ and $b_{m}$ by substituting the interior differentials 
with the combination of ghosts and we have 
\eqn\pcoD{
\delta(\iota_{\tilde v}) = 
\delta\Big(v^{\a} w_{\a}\Big) 
+ i \delta'\Big(v^{\a} w_{\a}\Big) v^{m}b_{m} 
- {1\over 2} \delta''\Big(v^{\a} w_{\a}\Big) v^{m} v^{n} b_{m} b_{n} 
- {i\over 6} \delta'''\Big(v^{\a} w_{\a}\Big) v^3 b^3 \,, 
}
where $v^{3} = \e_{mnr} v^{m}v^{n}v^{r}$ and 
$b^{3} = \e_{mnr} b^{m} b^{n} b^{r}$. 
A fundamental property used in the next sections is 
\eqn\pcoE{
\iota_{\tilde v} \delta(\iota_{\tilde v}) = 0\,.
}
It follows directly by the integral definition \pcoC. 
For an even vector field $v$ 
we identify $\delta(\iota_{v}) = \iota_{v}$ and 
equation \pcoE\ is obvious. 

Finally, 
we can define the Picture Changing Operator 
as follows
\eqn\pcoD{
Z_{\tilde v} = \Big[Q, \Theta(\iota_{\tilde v}) \Big] = 
\delta(\iota_{\tilde v}) {\cal L}_{\tilde v} + \delta'(\iota_{\tilde v}) 
\iota_{\tilde v^{2}} \,,
}
where 
\eqn\pcoE{\eqalign{
&\tilde v^{2} = 
\left( v^{\b} D_{\b} v^{\a} + v^{m} \p_{m} v^{\a} \right) d_{\a} + 
\left( v^{\b} D_{\b} v^{n} + v^{m} \p_{m} v^{n} \right) P_{n} \,, \cr
&\iota_{\tilde v^{2}} = 
\left( v^{\b} D_{\b} v^{\a} + v^{m} \p_{m} v^{\a} \right) w_{\a} + 
\left( v^{\b} D_{\b} v^{n} + v^{m} \p_{m} v^{n} \right) b_{n} \,,
}}

The general form of those operators is rather 
complicate, however one can choose the most 
convenient vector $\tilde v$ in order to simplify those 
operators. The amplitudes will not depend on the choice of 
the odd vector $\tilde v$ since
\eqn\varB{
\delta_{\tilde v^{\a}} Z_{\tilde v} = 
\Big[Q, (\delta_{\tilde v^{\a}} \iota_{\tilde v}) 
\delta( \iota_{\tilde v})\Big]\,.
}
Notice that, as opposed to $Z_{\tilde v}$, its 
variation is a BRST variation of a pseudoform. Therefore, 
by inserting the PCO in the amplitudes we are guaranteed that 
the latter are independent of the choice of the gauge parameters 
$\tilde v$. In the same way, the dependence of the PCO upon the 
time (the worldline coordinate) is also a BRST variation 
of a delta function which is an element of the space 
of the pseudoforms. This implies that the PCO are 
independent of the position on the worldline. 

The odd vector can be field-dependent or 
field-independent. For example, let us choose 
$\tilde v^{\a}_{1} =constant$ or a ghost-number one 
combination $ \tilde v^{\a}_{2} = B_{mn} (\g^{mn} \l)^{\a}$ 
where $B_{mn}$ is constant, we have
\eqn\simplePCO{
Z_{\tilde v_{1}} = \tilde v^{\a}_{1}(d + b^{p} \g_{p} \l)_{\a} 
\delta( \tilde v^{\a}_{1} w_{\a}) \,,
}
$$
Z_{\tilde v_{2}} = 
B_{mn} \l \g^{mn} (d + b^{p} \g_{p} \l)
\delta( B_{mn} \l \g^{mn} w) \,,
$$ 
where it can easily be seen that in the second operator, the 
generator of the Lorentz transformations 
$N^{mn} = \l \g^{mn}w$ appears. The gauge parameters 
$B_{mn}$ can be chosen in such a way that the no normal ordering 
is necessary to define $Z_{\tilde v_{2}}$. 
These PCO have been discussed 
and used in \BerkovitsPX\ and in \AnguelovaPG. 

Acting on the space of superforms on ${\cal M}$, 
the PCO $Z_{\tilde v_{1}}$ becomes 
\eqn\simplePCOA{
Z_{\tilde v_{1}} \Omega^{(p|q)} 
=
\Big[ 
\tilde 
v^{\a} (D + \l \g^{m} \partial_{c^{m}})_{\a} \delta(\tilde v^{\a}
\partial_{\l^{\a}})
\Big]  
\Omega^{(p|q)} \rightarrow \Omega^{(p|q-1)}\,.
 }
 The delta function $\delta(\tilde v^{\a} \partial_{\l^{\a}})$ 
 reduces the number of delta functions $\delta(\l^{\a})$ 
 present in the pseudoform since 
 $$
 \delta(\tilde v^{\a} \partial_{\l^{\a}}) \delta( f_{\a} \l^{\a}) = 
 \int_{-\infty}^{\infty} dt \, e^{i t \tilde v^{\a} \partial_{\l^{\a}} } \, 
 \delta( f_{\a} \l^{\a}) = \int_{-\infty}^{\infty} dt \,
 \delta\Big( f_{\a} ( \l^{\a} + i t v^{\a}) \Big) = - i { 1\over
 f_{\a} {v^{\a}}}\,.
$$
where $f_{\a}$ is a $\l$-independent parameter.\foot{In conformal 
field theory 
$$
\delta(\tilde v^{\a} w_{\a}(y))  
\delta( f_{\a} \l^{\a}(z)) = 
 \int_{0}^{\infty} dt \, 
 e^{i t \tilde v^{\a} w_{\a}(y)} \, 
 \delta( f_{\a} \l^{\a}(z)) = 
 $$
  $$
 \int_{0}^{\infty} dt \,
 \delta\Big( f_{\a} ( \l^{\a} + {i t \over (y-z)} v^{\a}) \Big) = - i { (y-z) 
 \over f_{\a} {v^{\a}}}\,.
 $$
 So, if $
 \Omega^{(0|1)} =
 i (f_{\a} \t^{a}) \delta(f_{\a} \l^{\a})$, then 
 we have that 
 $$\lim_{y\rightarrow z} 
 \Big(Z_{v_{1}}(y) \Omega^{(0|1)}(z)\Big) = 
 \lim_{y\rightarrow z} \Big[
 \tilde 
 v^{\a}_{1} (d + b^{m} \g_{m} \l)_{\a}(y) \delta(
 \tilde v^{\a}_{1} w_{\a}(y)) 
 \Omega^{(0|1)}(z) \Big] = 1 \,.
$$ 
 So, the vertex $\Omega^{(0|1)} =  i (f_{\a} \t^{a}) \delta(f_{\a} \l^{\a})$, 
 which is BRST invariant, is the inverse picture changing 
 operator. Notice that the gauge parameters $\tilde v^{\a}$ 
 and $f_{\a}$ cancel out. 
 } Notice that the operator $\delta(\tilde v^{\a} \p_{\l^{\a}})$ 
 is ill-defined on pseudoforms which carry a different 
 picture ($v^{\a} f_{\a} =0$). In that case, we define it 
 to vanish.  So, effectively the 
PCO reduce the number of delta functions in the vertex operator belonging to 
$\Omega^{(p|q)}$. 

The PCO $Z_{\tilde v}$ are designed to reabsorb the 
zero modes of $w_{\a}$ in the path integral measure. However, 
the number of zero modes is $n \times g$ where $n$ is the 
Grassmann dimension of the superspace ${\bf R}^{(m|n)}$ and 
$g$ is the genus of the Riemann surface. Therefore, we need 
to stack a corresponding number of PCO. This can be done 
by choosing a basis of abelian differential $\omega^{\a}_{i}$ 
where $\a = 1,\dots, n$ and $i = 1,\dots,g$ defined 
such that 
\eqn\abA{
\oint_{a_{i}} \omega^{\a}_{j} = \delta_{ij} v^{\a}
}
where $a_{i}$ is the $a$-cycle on the Riemann surface and 
$v^{\a}$ is a constant (or depending on the zero modes 
$\t^{\a}$, $\l^{\a}$ and $x^{m}$).  
Using this basis, we can finally construct the operator 
\eqn\abB{
\prod_{i=1}^{g}\prod_{\a_{i} =1}^{n} Z_{\omega^{\a}_{i}}
}
which maps the cohomology $H$ onto the cohomology 
group a different number of pictures. 
Also in the present case we conjecture that the number of 
zero modes for $w_{\a}$ and $b_{m}$ is equal to number of 
differential forms on a genus $g$ Riemann surface. However, in 
the particle limit the antighost fields loose their conformal weight. 
We assume however that the number of zero modes remains invariant. 
(An analysis of higher loop expansion for superparticle model 
was performed in \GrassiXR). The problem to derive this number 
of zero modes has to be ascribed to the non-manifold nature of particle 
graph. Those graphs are embedded into string theory graph and 
the conformal weights of the fields play an important role.

Let us now consider the odd form of $\Omega^{(1|0)}$:
$\tilde f =  f_{\a} \l^{\a} + f_{m} c^{m}$ where $f_{\a}$ and 
$f_{m}$ are commuting and anticommuting, respectively.
These components can also be field dependent and the 
total parity of the odd form is 0. We 
define the exterior product as usual:
\eqn\eA{
e_{\tilde f}:\Omega^{(n|p)} \rightarrow \Omega^{(n+1|p)}\,, 
~~~~~
e_{\tilde f} \omega = \tilde f \wedge \omega\,.
}
This operation, together with $\iota_{\tilde v}$ for an odd vector
$\tilde v$, satisfies the following Clifford algebra: 
\eqn\eB{
\Big[e_{\tilde f}, e_{\tilde g} \Big] =0\,, ~~~~ 
\Big[e_{\tilde f}, \iota_{\tilde v} \Big] = \tilde f(\tilde v) \equiv 
f_{\a} v^{\a} + f_{m} v^{m}\,.
}
Again, since the operation $e_{\tilde f}$ is an even 
operation, we can define Dirac delta functions of this operator. 
In  particular we can construct an inverse operation 
to $Z_{\tilde v}$ as follows: given an anticommuting 
superfield $f = f(x,\t)$ on the supermanifold, its 
differential $d f = \{Q, f\} \equiv \tilde f \in \Omega^{(1|0)}$ 
is an odd differential form and we define the inverse PCO 
with 
\eqn\inA{
Y_{f} = f(x,\t) \delta( e_{ \{Q, f\} })\,.
} 
It is obvious that $Y_{f}$ is BRST invariant, in fact action 
with $Q$ on $Y_{f}$, one has $\{Q, Y_{f}\} = 
\{Q, f\} \delta(e_{\{Q, f\}}) =0$ because of the property 
of delta functions. In addition, any change of the 
function $f$ is BRST exact
\eqn\inB{
\delta Y_{f} = \delta f  \delta( e_{ \{Q, f\} }) - 
f \{Q, \delta f\} \delta' ( e_{ \{Q, f\} })  =
\Big[ Q, f \delta f  \delta( e_{ \{Q, f\} }) \Big]\,. 
}
It is therefore convenient to choose very simple representation 
for $Y_{f}$, such as, for example those 
related to the choices 
$f = f_{\a} \t^{\a}$
and with  $f_{\a} = constant$ or 
field dependent $f_{\a} = w_{\b} C^{\b}_{~~\a}$
(with $C^{\b}_{\a} = constant$)
\eqn\inC{
Y_{f_{1}} = f_{\a} \t^{\a} \delta(e_{f_{\a} \l^{\a}})
\,.
}
$$
Y_{f_{2}} = 
\Big( w_{\b} C^{\b}_{~~\a} \t^{\a} \Big)
\delta(
e_{  [(d + b^{p} \g_{p} \l)_{\b} C^{\b}_{~~\a} \t^{\b} + 
w_{\b} C^{\b}_{~~\a} \l^{\a} ] })
\,.
$$
In the following, we will just write $\tilde f$ instead of $e_{\tilde
f}$ in the delta function since there is no ambiguity. 

Given a set of anticommuting 
functions $f_{i}$ where $i=1, \dots, n$ where 
$n$ is the odd dimension of the superspace ${\bf R}^{m|n}$ and 
$n$ is the maximal number of delta functions of a given vertex, we 
can define the PCO 
\eqn\inD{
Y_{\{f_{i}\}_{i}} = \prod_{i=1}^{n} f_{i} \delta(\{Q, f_{i}\})
}
which has the same properties of $Y_{f}$, but this 
maps 
the cohomology $H^{(m|0)}(Q)$ into $H^{(m|n)}(Q)$, the cohomology 
group with the maximal number of delta functions.  

\subsec{Changing the Number of Delta Functions}

To show how the PCO work, we consider several 
examples. 
First, we can go back again to the simplest case of ${\bf R}^{1|1}$
with the de Rham differential. We observed after \derhama\ that there is a 1-1
correspondence between $H^{*|0}$ and $H^{*|1}$. We can now exhibit the
isomorphism as a PCO. In this case there is only one constant odd vector field,
namely $\tilde v = {\p \over \p\t}$, and the corresponding
picture-lowering operator is $Z = \delta(\iota_{\tilde v}) {\cal
L}_{\tilde v}$ and acts as $\theta \delta(\lambda) \mapsto 1$. The
inverse operation is evidently $Y = \theta \delta(\l)$, which 
corresponds to \inA\ with $f = \theta$. The two operators $Z$ and
$Y_f$ are in this case inverses on the whole complex $\Omega^{*|*}$. A
different choice of $f$ would give an equivalent $Y$, which would in
general fail
to be an exact inverse to $Z$, but would still be an inverse in the
cohomology.  

We consider the vertex operator 
$U^{(1|0)} = \l^{\a} A_{\a} - c^{m} A_{m}$ 
described in sec. 3.1, and we can act with the PCO 
$$Y_{\{f^{(\b)}\}} = \prod_{\b =1}^{2} 
(f^{(\b)}_{\a} \t^{\a})  \delta(f^{(\b)}_{\a} \l^{\a}) = 
\t^{2} \delta^{2}(\l^{\a})$$ (where we normalize the function $f$ 
such that $f^{(\g)}_{\a} f^{(\d)}_{\b} \e^{\a\b} =\e^{\g\d}$) to have
\eqn\inE{
U^{(1|2)} =  
Y_{\{f^{(\b)}\}} U^{(1|0)} =  
\t^{2} \delta^{2}(\l) (\l^{\a} A_{\a} - c^{m} A_{m})  = 
- c^{m} (A'_{m}) \delta^{2}(\l)
} 
where $A'_{m} = A_{m} \t^{2} = a_{m}(x) \t^{2}$. It is easy to 
check that $U^{(1|2)}$ is in the cohomology $H^{(1|2)}(Q)$ if 
$a_{m}$ is a flat connection. Since $U^{(1|0)}$ was in the 
cohomology $H^{(1|0)}(Q)$, one can easily show that 
the new vertex operator $U^{(1|2)}$ depends upon the 
choice of the parameters $f_{\a}$ in the PCO only through 
BRST exact terms which do not affect the amplitudes.  
We can also compute the transformed vertex operator 
$U^{(2|2)}$ as follows
\eqn\inG{
U^{(2|2)} =  
Y_{\{f^{(\b)}\}} U^{(2|0)} =
\t^{2} \delta^{2}(\l) (\l^{\a} \l^{\b} A^{*}_{\a\b} - 
c^{m} \l^{\b} A^{*}_{m \b} 
c^{m} c^{p} A^{*}_{[mp]}
)   
}
$$
 = - c^{m}c^{p} (A^{*}_{[mp]} \t^{2}) \delta^{2}(\l)
$$
which is again in the cohomology $H^{(2|2)}$.  

In the same way, we can consider the vertex operator in the 
zero momentum cohomology $U^{(3|0)}$ given in \zmc\ 
and we can construct the vertex $U^{(3|2)}$ with 2 delta functions as follows
\eqn\inF{
U^{(3|2)} = Y_{\{f^{(\b)}\}} U^{(3|0)} = 
c^{m} c^{n} c^{r} \e_{mnr} \t^{2} \delta^{2}(\l)
}
which is the top form of ${\bf R}^{(3|2)}$ and it can
be integrated on this space. Again, the dependence on 
$f_{\a}^{(\b)}$ is only through BRST exact terms and $U^{(3|2)}$ is 
representative of the cohomology group $H^{(3|2)}(Q)$ at zero momentum. 

It is also very interesting to use the PCO $Z_{\tilde v}$ 
to lower the number of delta functions in  the vertex operators. For that purpose, 
we consider the simplest PCO 
$Z_{\tilde v} = \tilde v^{\a} (d + b_{m} \g^{m} \l)_{\a} \delta(\tilde
v^{\a} w_{\a})$  
where $\tilde v^{\a}$ is a constant parameter and we 
act on the vertex operator $U^{(1|2)}$ derived above \inE. 
We have 
\eqn\pA{
Z_{\tilde v} U^{(1|2)} = 
\tilde v^{\a} (d + b_{m} \g^{m}\l)_{\a} \delta(\tilde v^{\a} w_{\a})
\Big[c^{m} \tilde A_{m} 
\delta(f_{\a} \l^{\a}) \delta(f_{\perp} \l^{\a}) \Big] 
}
$$
= 
\tilde v^{\a} (D + \l\g^{m}\partial_{c^{m}})_{\a} 
\delta(\tilde v^{\a} \p_{\l_{\a}})
\Big[c^{m} \tilde A_{m} 
\delta(f_{\a} \l^{\a}) \delta(f_{\perp} \l^{\a}) \Big] 
$$
$$
 = \Big( c^{m} \tilde v^{\a} D_{\a} \tilde A_{m} + 
(\l \g^{m} \tilde v) \tilde A_{m}\Big) \delta(f_{\perp,\a} \l^{\a})
$$
where we choose to decompose the product 
$\delta^{2}(\l)$ along two orthogonal spinors 
$F^{i}_{\a} = (f_{\a}, f_{\perp, \a})$ such that 
$\e^{\a\b} F^{i}_{\a} F^{j}_{\b} = \e^{ij}$. In addition, 
we also choose $\tilde v^{\a} f_{\a} =1$ which is 
not essential and will not change our conclusions. This however 
implies, as seen above, that 
$\delta(\tilde v^{\a} w_{\a}) \delta(f_{\a} \l^{\a}) = 1$.   
\foot{To justify the above 
manipulations is better to use CFT techniques and 
we have $U^{(1|1)} = 
\lim_{x \rightarrow y} :Z_{\tilde v}(x) U^{(1|2)}(y): \,.$ 
The normal ordering removes 
all poles from the contraction of the PCO with vertex operator and 
the limit removes the zeros of the form $(z-w)^{n}$ leaving only 
the finite and non vanishing parts.  
} 

We can repeat the operation with an orthogonal 
PCO $Z_{\tilde v^{\a}_{\perp}} = 
\tilde v^{\a}_{\perp} 
(d + b_{m} \g^{m} \l)_{\a} \delta(\tilde v^{\a}_{\perp} w_{\a})$. 
This gives 
\eqn\pB{
U^{(1|0)} = Z_{\tilde v_{\perp}} Z_{\tilde v} U^{(1|2)} = 
\tilde v^{\a} \tilde v^{\b}_{\perp} 
\Big[  c^{m} D_{\a} D_{\b} \tilde A_{m} + 
(\g^{m} \l)_{(\a} D_{\b)} \tilde A_{m} \Big]
} 
Inserting the definition of $\tilde A_{m}$, one gets the 
original vertex operator \brsAA\ $U(1|0) = \l^{\a}A_{\a} + c^{m} A_{m}$ 
with 
\eqn\bC{
A_{\a}(x,\t) = 
\Big((\g^{m} \t)_{\a} + (\tilde v\g^{m}\tilde v_{\perp}) \t_{\a}\Big)
a_{m}(x)\,, } 
$$
A_{m}(x, \t) = a_{m}(x) + \t^{2} (\tilde v\g^{p} \tilde v_{\perp})
\p_{m} a_{p}(x)\,.  
$$
These superfields differ from the orignal solution found in 
sec. 3.1 by a choice of gauge. Indeed, it can be shown that 
the vertex operator satisfied the Wess-Zumino-like gauge condition  
$$
\Big(\t^{\a} - (\tilde v\g^{m} \tilde v_{\perp}) (\g_{m} \t)^{\a}\Big)
A_{\a} = 0.  
$$
This is a Lorentz-transformed Wess-Zumino-like gauge $\t^{\a}
N_{\a}^{~\b} A_{\b} = 0$ where $N_{\a}^{~\b}$ is the infinitesimal
Lorentz  
transformation whose parameter is 
$\omega_{mn} = e_{mnp} (\tilde v\g^{p} \tilde v_{\perp})$. 
This is not surprising 
since to define the PCO we had to choose a gauge given 
by the constant parameters $\tilde v^{\a}$ and $\tilde v^{\a}_{\perp}$.

\subsec{Genus $g$ differential forms and anomalies}

With the geometric theory of integration at our disposal, we only need
one more ingredient before we can proceed to present the computation
of the correlators. As we have already remarked, at tree-level there
is 
a perfect identification of the superforms on the target manifold with
the vertex operators that depend only on the zero modes of the scalar
fields, in our case $x^{m}, \t^{\a}, \l^{a}$ and 
$c^{m}$. We can denote this as tree-level forms. As soon as we want to
consider higher-genus Riemann surfaces, 
we have to take into account the presence of the conjugate momenta,
resp. $P_{m},  p_{\a},w_{\a}$ and $b_{m}$;
these are fields of conformal weight 1 and have $g$ zero modes on a
surface $\Sigma$ of genus $g$. Making a 
choice of a symplectic base $a_i,b_i, i =1, \ldots, g$ for the first
homology group of $\Sigma$, the zero-modes can be conveniently
described as the integrals $\oint_{a_{i}} w_{\a z}(z) dz$ (and
similarly for the other fields). 

Since in the particle limit the Riemann surface degenerates into a
graph and one cannot define a notion of conformal weight, and even the
counting of zero modes becomes ill-defined, we have to consider this
issue of higher-loops in the full string theory. Once we have written
the prescriptions, they can be used also for the particle. 

It is then convenient to use a conformal field theory
formalism. We can introduce conserved currents that assign quantum
numbers to the ghosts, and the number of insertions that are necessary
to absorb the zero modes are counted by the anomaly in their OPE with
the stress energy tensor. In the usual topological sigma model there
is only one such current, 
associated to the ghost number. In our case we will need another one
associated to the picture. There is a certain freedom in defining
these currents, simply because any linear combination of them would
also be conserved. We will make the following choice:
\eqn\ghostcurrents{
J_{gh} = c^{m} b_{m} - \l^\a w_\a \,, ~~~~~ 
J_{pi}  = - \Theta(w_\a) \partial \Theta(\lambda^\a) \,, 
} 
These expressions need some clarification. In order to compute their
anomalies and their assignment of charges to the various fields, it is
more 
convenient to use a bosonized form for the commuting ghosts, even
though this breaks the manifest Lorentz invariance. Then we can use
the following dictionary (see e.g. 
\lref\VerlindeKW{
E.~Verlinde and H.~Verlinde,
Nucl.\ Phys.\ B {\bf 288}, 357 (1987).
}
\VerlindeKW ): for each pair of conjugate
fields ($\l,w$) (in our case there are 2 such pairs),  
\eqn\bosonized{\eqalign{
\l = \eta e^{\phi} \,, & \quad w = \p \xi e^{-\phi} \,, \cr
\delta(\l) = e^{-\phi} \,, & \quad \delta(w) = e^{\phi} \,, \cr
\p \Theta(\l) = \p \l \delta(\l) = \eta \,, & \quad \Theta(w) = \xi \, \cr 
\l w = \p \phi \,. &  \,
}}
Here $\xi,\eta$ form as usual an anticommuting system of spin $(0,1)$,
and $\phi$ is a chiral boson with background charge $-1/2$, so that the
exponentials in the first line of \bosonized\ have weights $-1$ and 0
respectively. 

Then one can see easily that $\l,w$ have ghost number $\pm 1$ and 
picture $\mp 1$, while $\delta(\l),\delta(w)$ have ghost number $\mp 1$ and no
picture. This definition of the ghost current is in
agreement with the requirement that its anomaly should count the
number of zero modes of the fields charged under it. In fact, an
insertion of $\delta(\lambda)$ can be used to reabsorb a zero mode of
$\lambda$, and so it carries the opposite ghost number. 
Using the bosonization rules \bosonized\ it is straightforward to compute the
anomalies of the currents:
\eqn\anomalies{
c_{gh} = -3 + 2 = -1 \, , \quad c_{pi} = -2 \, .
}
This result will be important when we discuss the
prescription for the computation of amplitudes. At tree level, an operator that 
saturates the anomaly is $c \l^2 \t^2 \delta^2(\l)$, and we have already seen 
that with the measure given by \measu\ , this operator has a non-vanishing vev. 
So the selection rules given by the anomalies are consistent with the tree level 
prescription we have found, and moreover they enable us to extend the prescription 
to higher loops. One has to recall that the anomaly is proportional to the Euler 
character of the Riemann surface, and at genus $g$ it is given by $c(1-g)$, where 
$c$ is the coefficient computed by the OPE with the stress-energy tensor. 

We note also that -- as in the RNS formalism -- the BRST charge $Q$ is 
filtered with the respect to this number as follows $Q = Q_{0}  + Q_{-1} + Q_{-2}$ 
where $Q_{0} = c_{m}  P^{m}, Q_{-1} = \l^{\a} d_{\a}$ and 
$Q_{-2} = {1\over 2}b_{m} \l \g^{m} \l$. 
In the same way the vertex operators are filtered according to this number. 
 
Whereas for tree-level forms we can identify the measure with an element of the 
zero-momentum cohomology of top degree, which usually turns out to be unique, 
this is not the case for higher loop forms. For instance, let us consider the 
case of 1-loop forms. By using the unitary transformation 
\actH\ , the BRST operator becomes simply $Q = \l^\a
\hat p_\a - {\hat c}^m P_m$ and we are effectively computing the de
Rham cohomology in the space of 1-loop forms on ${\cal M}$. This can
be written as  
\eqn\loopcohom{
\prod_{m=1}^3 \{1,\, {\hat c}^m\} \otimes \prod_{n=1}^3 \{1, \, b^n\} \otimes 
\prod_{\a=1}^2 \{1,\, \t^\a \delta(\l^\a)\} \otimes 
\prod_{\b=1}^2 \{1, \, \hat p_\b \, \delta(\hat w_\b) \} \,.
} 
We are particularly interested in the space $H^{0|0}$ that should
be relevant for the 1-loop integration measure, since all the anomalies vanish 
at genus 1. But it is clear from \loopcohom\ that this space is not 1-dimensional. 
There is however a distinguished element, which contains the zero modes of 
all the fields, and is the only class 
$\omega \in H^{3|0}(x, \hat c) \otimes H^{-3|0}(b,P) \otimes
H^{0|2}(\t,\l) \otimes H^{0|-2}(\hat p, \hat w)$. Explicitly, $\omega
= c^3 \, b^3 \, \t^2 \, p^2 \, \delta^2(\l) \, \delta^2(w)$. Note that
$\omega$ has the same form in term of the transformed and the original
fields. This cohomology class is the natural volume form on ${\cal
M}\oplus \hat{\cal M}$, and determines a pairing $<,> : \, H^{p|q}
\times H^{-p|-q} \to {\bf C}$. The space $H^{0|0}$ is dual to itself
under this pairing, and $<1, \omega > = 1$.  It is natural to consider the factors 
in $\omega$ depending on the ghosts as related to the measure for the tree-level 
fields; then the antighost introduce additional zero modes that make the total 
anomaly vanish and require the introduction of the PCO in the correlators. 
Analogous statements hold for higher loops, as will be illustrated in the next section.


\newsec{Multiloop multilegged amplitudes}

In this section, we clarify completely the prescription 
for multiloop computation in the 
present model. The prescription is 
obtained by mapping the Riemann surface into 
graphs. We separately discuss the tree level (which has been 
discussed in pevious sections), one loop and finally for $g >1$. 
This is needed in order to fix the isometries of tree level and 
one loop. 

\subsec{$g=0$, N-point functions}

We start to give the prescription of tree level multilegged 
amplitude. In the present situation, there is no zero modes for 
the conjugated momenta $d_{\a},w_{\a}$ and $b_{m}$. 
It is easy to see that, based on the previous considerations, the 
amplitude is given by
\eqn\zeroN{
{\cal A}^{N}_{g=0} \sim 
\int \Big(d^{3}x_{0} d^{3}P \hat{\cal D}x \Big) d^{2}\t d^{2}\l d^{3}c
\Big( (\not\!c \partial_{\l})\cdot (\not\!c \partial_{\l}) \delta^{2}(\l) 
\Big) \times }
$$
\times
\prod_{i=1}^{3}U_{i}^{(1)} \prod_{j=4}^{N} \int d\tau_{j} 
V^{(0)}_{j}(\tau_{j}) \,,
$$
 where $x_{0}$ is the center of mass and ${\cal D}x$ is the 
measure for non zero modes and 
\eqn\zeroNA{
\langle x^{m}(\tau_{1}) x^{n}(\tau_{2}) \rangle_{g=0} = 
\int \hat{\cal D}x \Big(x^{m}(\tau_{1}) x^{n}(\tau_{2})\Big) =
\eta^{mn}|\tau_{1} - \tau_{2}|\,.
}
The insertion of 3 vertex operators $U^{(1)}$  
with ghost number one is needed in order to 
fix the isometries at tree level (in string theory 
this corresponds to fixing the M\"obious group $SL(2,{\bf R})$ 
of the disk). In the case of the particle limit, one of the vertex $U^{(1)}$ 
is situated at the beginning of the worldline ($T = -\infty$) one 
at the other end and the third in the middle for a generic $T$ 
(this analysis can be found in 
\lref\StrasslerZR{
M.~J.~Strassler,
Nucl.\ Phys.\ B {\bf 385}, 145 (1992)
[arXiv:hep-ph/9205205].
}
\lref\SchubertHE{
C.~Schubert,
Phys.\ Rept.\  {\bf 355}, 73 (2001)
[arXiv:hep-th/0101036].
} \refs{\StrasslerZR,\SchubertHE}). The insertion of integrated 
operators is needed to have $N$-point amplitudes and they are given 
by $\int d\tau V^{(0)}(\tau) = \int d\tau \dot x^{m}(\tau) A_{m}(x(\tau))$. 
The form of the connection $A_{m}$ depends on the background chosen. 
Example of these amplitudes are already given in the previous sections.  

Regarding the saturation of the anomalies, we can easily see that 
if we consider the combination 
$ [{\cal D}c {\cal D}\l]_{(-1|2)} = d^{2}\l d^{3}c
(\not\!c\partial_{\l})\cdot (\not\!c \partial_{\l})$ 
as part of the measure (which defines the vacuum of the Hilbert space), 
it has ghost anomaly equal to $-1$ which is the 
ghost charge of the vacuum while it has picture number equal to 
$2$ which is again  what dictates the picture number anomaly. 
On the other side the insertions, namely 
$\prod_{i=1}^{3} U^{(1)}_{i}$ and $\delta^{2}(\l)$, 
have ghost number $+1$ 
which saturate the ghost anomaly of the vacuum and out to 
the three vertex operators $U_{i}^{(1)}$ we have to select the 
picture $-2$ part. That leads to a non-vanishing amplitude.\foot{In pure 
spinor approach the measure for the zero modes of 
open superstrings is given by  
$[{\cal D \l}]_{(8|3)} = d\l^{\a_{1}}\wedge \dots \wedge d\l^{\a_{11}} 
\e_{\a_{1} \dots \a_{16}} 
(\g_{m} \g_{n} \g_{r} \g^{mnr})^{[\a_{12} \dots \a_{16}] (\b_{1} \dots \b_{3})} 
{\partial \over \partial \l^{\b_{1}}} 
\dots 
{\partial \over \partial \l^{\b_{3}}}$. This measure has 
ghost number +8 as prescribed by the ghost anomaly $ T J \sim -8/(z-w)^{3} + \dots$ (see \BerkovitsPX\ for further details) and has picture +3, and 
this is prescribed by a similar operator as in the previous section. 
} 
 
\subsec{$g=1$, N-point functions} 

The next step is to consider one loop amplitudes. Here the 
presence of zero modes for the conjugated variables 
$d_{\a}, w_{\a}$ and $b_{m}$ is important. For one-loop 
one has to integrate over the modulus of the torus $T$. 
To this modulus, there is a corresponding antighost field 
denoted $B_{T}$. 
The one-loop N-point function is given by
\eqn\zeroN{
{\cal A}^{N}_{g=1} \sim 
\int_{0}^{1} {dT } 
\int \Big( d^{3}x_{0} d^{3}P \hat{\cal D}x\Big) 
d^{2}\t  d^{2}d d^{2}\l d^{2}w d^{3}c  d^{3}b
\Big( (\not\!c \partial_{\l})\cdot (\not\!c \partial_{\l}) \delta^{2}(\l) 
\Big) \times }
$$
\times
B_{T} \prod_{k=2}^{N} \int_{0}^{T} d\tau_{k} B_{k} 
\prod_{i=1}^{2} Z_{v_{i}} U^{(1)}(T) 
\prod_{j=2}^{N} U^{(1)}_{j}(\tau_{j})\,,
$$ 
where $B = b_{n} \dot x^{n}$ and 
$$Z_{v_{i}} = \{Q, \Theta(v^{\a}_{i} w_{\a})\} = 
v^{\a}_{i}\Big( d - b_{n} \g^{n}\l\Big)_{\a} \delta(v^{\a}_{i}w_{\a})\,.$$ 
The parameters $\tau_{j}$ are the Schwinger parameters. 
The counting of zero modes goes as follows: 

{\it 1)} there are 3 zero modes for $c^{m}$ and 3 zero modes for
$b_{m}$. Notice  
that $b_{m}$ behave like abelian differentials of string theory, 
(we have to recall here that the 1-loop graph is indeed a manifold 
and the abelian differentials on it are well defined. There 
is one modulus for the torus for each vertex insertion. For 
higher loops the counting of moduli cannot be done in the same way 
since higher loop graphs are not manifolds.

{\it 2)} there are 2 zero modes for $\t^{\a}$ and 2 zero modes of 
$d_{\a}$. 
They are fermionic and they have to be saturated in order 
to get a non-vanishing amplitude, 

{\it 3)} there are 
2 zero modes of $\l^{\a}$ and 2 zero modes for $w_{\a}$. They 
are bosonic and the delta functions $\delta^{2}(\l)$ and $\delta^{2}(w)$ have 
to be inserted. The first delta functions are inserted by means of the 
picture lowering operator 
$\Big( (\not\!c \partial_{\l})\cdot (\not\!c \partial_{\l})
\delta^{2}(\l) \Big)$ 
-- which 
is BRST invariant\foot{The BRST variation of $c^{m}$ produces $\l \g^{m} \l$, 
and therefore using the two derivatives appearing on the delta function, it 
is easy to see that the BRST variation of this monomial is zero. } -- there 
are two derivatives on delta functions which select those products of vertices 
that contain at least two powers of $\l^{\a}$. The second type 
of delta functions $\delta^{2}(w)$ is contained in the picture changing 
operators $Z_{v}$. They depend on the gauge parameters $v^{m}_{i}$ where 
$i=1,2$ and they are chosen in such a way that no normal ordering is 
needed for the expression of $Z_{v}$. This dependence is BRST 
exact $$\delta_{v^{\a}_{i}} Z_{v} = \{Q, \delta_{i}^{j} w_{\a} 
\delta(v^{\a}_{j} w_{\a}) \}\,.$$ 

4) The total ghost number of the amplitude as it can be checked by 
$$
- 2_{\delta^{2}(\l)}  - (N+1)_{(B_{k})} + 2_{(Z_{v})} + (N +1)_{(U^{(1)})} =0$$
where in bracket we inserted the sources of the ghost charge. 

5) We also have to take into account that the picture 
should be saturated. Indeed we have
$$ -2_{(\delta^{2}(\l))} + 2_{(Z_{v})}  =0\,.$$

As an example, we consider one-loop, three point function. 
\eqn\oneloopthreepoint{
{\cal A}^{3}_{g=1} \sim 
\int_{0}^{\infty} {dT \over T} 
\int \Big( d^{3}x_{0} d^{3}P \hat{\cal D}x\Big) 
d^{2}\t  d^{2}d d^{2}\l d^{2}w d^{3}c  d^{3}b
\Big( (\not\!c \partial_{\l})\cdot (\not\!c \partial_{\l}) \delta^{2}(\l) 
\Big) \times }
$$
(b_{m} \dot x^{m})(T) 
\int_{0}^{T} d\tau_{1} (b_{m} \dot x^{m})(\tau_{1}) 
\int_{0}^{T} d\tau_{2} (b_{m} \dot x^{m})(\tau_{2}) \times
$$
$$ 
v^{\a}_{1} \Big(d - b_{m} (\g^{m} \l)\Big)_{\a} 
v^{\b}_{2} \Big(d - b_{m} (\g^{m} \l)\Big)_{\b} 
\delta(v^{\a}_{1} w_{\a}) \delta(v^{\b}_{2} w_{\b}) 
\times
$$
$$
(\l^{\a}A_{\a} - c^{m}A_{m})
(\l^{\a}A_{\a} - c^{m}A_{m})
(\l^{\a}A_{\a} - c^{m}A_{m})\,.
$$
It is easy to check that indeed all the zero modes 
are saturated and therefore this amplitude does not vanish. To evaluate it, 
one needs to compute all the integration on zero modes and 
finally the contraction of $x$'s is performed with the 
propagator
\eqn\oneloopprop{
\langle x^{m}(\tau_{1}) x^{n}(\tau_{2}) \rangle_{g=1} 
= \eta^{mn} \Big( |\tau_{1} -\tau_{2}| - 
{(\tau_{1} -\tau_{2})^{2} \over T} \Big)\,.
}

To define the partition function at one loop, one 
has to insert powers of the ghost current $J = b_{m}c^{m} + w_{\a} \l^{\a}$, 
but it  vanishes as it can be seen by just counting the zero modes. This is 
equivalent to the vanishing of the partition function 
for Green-Schwarz superstring in the light-cone gauge on the 
flat background. 

\subsec{$g > 1$, N-point functions}

In this last section, we consider the multiloop when 
$g>1$ and we do not have to take  into account any isometry of the 
graph. We have to notice that by counting the internal lines $I$
of a graph with $V$ internal vertices at $g$ loops, one has 
$g -1 =  I -V$. Since the vertices are trivalent $ 3 V = 2 I + N$ where 
$N$ is the external insertions, we have that $I = 3( g-1) + N$ which 
gives the number of internal lines to whom we assign a modulus each. 

By respecting the ghost number and zero modes saturation, we 
propose the prescription
\eqn\gN{
{\cal A}^{N}_{g} \sim \sum_{h=1}^{3g -3 +N} \int d m_{h}
\int d^{3}x_{0} d^{3}P \hat{\cal D}x 
d^{2}\t d^{2}\l d^{3}c 
\prod_{i=1}^{2g}d^{2}w_{i} d^{2}d_{i} \prod_{j=1}^{3g} d^{3}b_{j}
\times }
$$
\times
\Big( (\not\!c \partial_{\l})\cdot (\not\!c \partial_{\l}) \delta^{2}(\l) 
\Big) 
\prod_{k=1}^{3(g-1) + N} \int d\tau_{k} \mu_{k}(\tau_{k}) B_{k} 
\prod_{l=1}^{2g + N} Z_{v_{l}} \prod_{m=1}^{N} U^{(1)}_{j}(\tau_{j}) = \,, 
$$ 
$$
\sim \sum_{h=1}^{3g -3 +N} \int d m_{h}
\Big\langle
\Big( (\not\!c \partial_{\l})\cdot (\not\!c \partial_{\l}) \delta^{2}(\l) 
\Big) 
\prod_{k=1}^{3(g-1) + N} \int d\tau_{k}  \mu_{k}(\tau_{k}) B_{k} 
\prod_{l=1}^{2g + N} Z_{v_{l}} \prod_{m=1}^{N} U^{(1)}_{j}(\tau_{j})
\Big\rangle
\,. 
$$ 
In the present case, we notice that we add the contribution 
from the $3g$ zero modes of $b_{m}$ and 
$2g$ zero modes of $d_{\a}$ and of $w_{\a}$. They are 
saturated by the presence of $B$'s and by the picture changing operators. 
The counting of ghost number and picture number goes as follows
$$
- 2_{(\delta^{2}(\l))}  - (3(g-1) + N)_{(B_{k})} + (2g +N)_{(Z_{v})} +
N_{(U^{(1)})} =- (g-1) + N$$ 
$$ 
-2_{(\delta^{2}(\l))} + (2g +N)_{(Z_{v})} = 2 (g-1) + N\,.
$$
The last expression counts the number of delta functions 
for $\l$ and $w$ and the number is exaclty matched by the 
number of integration variables $d^{2}\l$ and $d^{2}w$. 

We notice therefore that in the case of a given number of possible pictures, 
let us say, $n$ different pictures, the counting of the insertions $Z_{v}$ and 
of the delta functions $\delta(\l)$ should be changed to $n$, 
namely the insertions should be replaced by 
$$
H(c, \partial_{\l}) \delta^{n}(\l) \prod_{l=1}^{n \, g + N} Z_{v_{l}}
$$
which conpensate the picture charge of the vacuum $n(g-1) + N$. 
\foot{
For example in the pure spinor approach, the number of independent 
components is computed by solving the constraint 
$\l^{\a} \g^{m}_{\a\b} \l^{\b} =0$, where $\l^{\a}$ is a Weyl spinor
in 10d and  
$\g^{m}_{\a\b}$ are symmetric real matrices of $Spin(10)$. This yields
11 components  
for $\l^{\a}$ and the insertion is given by 
$$
\Big(\t^{\a_{1}} \dots \t^{\a_{11}} \e_{\a_{1} \dots \a_{16}} 
\g_{m}^{\a_{12} \b_{1}} \g_{n}^{\a_{13} \b_{2}} \g_{p}^{\a_{14} \b_{3}} 
\g^{mnp, \a_{15} \a_{16}} \partial_{\l_{\b_{1}}}
\partial_{\l_{\b_{2}}} \partial_{\l_{\b_{3}}} 
\delta^{11}(\l)\Big) \prod_{l=1}^{11 \, g + N} Z_{v_{l}}
$$
where the term in the bracket absorbs the zero modes of $\l^{\a}$ and 
the PCO $Z_{v}$ absorbs the zero modes of $w_{\a}$. The total 
number of delta functions absorbed is $11 (g-1) + N$.} 

It is clear from this analysis that several amplitudes are vanishing. 
However, we have to recall that if the computation is done 
in topological field theory there some terms coming from the curvature 
terms of the action. This phenomena is illustrated in 
\lref\BCOV{
M.~Bershadsky, S.~Cecotti, H.~Ooguri and C.~Vafa,
Commun.\ Math.\ Phys.\  {\bf 165}, 311 (1994)
[arXiv:hep-th/9309140].
} \BCOV.  

\subsec{Derivation of \gN}

In order to derive the formulat \gN\ it is necessary to couple the theory 
to an extended topological system on the worldsheet (or on the worldline). 

There are essentially two type of zero modes 
relevant for our derivation: the zero modes and moduli of the 
Riemann surface and the supermoduli associated to commuting ghosts $\l^{\a}, w_{\a}$. 
The first type of zero modes are parametrized by suitable combinations of 
the 
ghost fields $(c^{m},b_{m})$, while the second by the 
ghosts $(\l^{\a}, w_{\a})$. In order to derive the formula \gN, we 
introduce four new type of ghost fields $(b',c')$, $(\b',\g')$ 
(with conformal spin (2,-1) and with fermionic and bosonic statistics,
respectively) and  
$(\b''_{i}, \g''^{i})$, $(\xi''_{i}, \eta''^{i})$ where 
$i =1,2$ (this number 
depends on the number of independent pictures) with conformal (1,0)  with 
bosonic and fermionic statistics. This set of fields resembles the 
set of ghost of a fermionic string with an extended supersymmetry, 
whose conformal central charge vanishes 
since these new fields always appear in quartets. On the other side, 
the anomaly of the ghost currents 
$$J_{gh} = b'c' + 2 \b' \g'\,, ~~~~~~~
J_{pi} = \b''_{\a} \g''^{\a} + 2 \xi''_{\a} \eta''^{\a}$$ 
are $-3$ and $- n$, respectively, where $n$ is the number of
independent components  
of $\l^{\a}$. 

Let us first derive the insertions of the $B = (b_{m} P^{m})$ 
of the formula \gN. 
They are related to the moduli of the Riemann surface and therefore 
it is useful to introduce a new BRST charge defined as follows
\eqn\gNN{
Q_{new} = c' P^{2} + \g' ( b' - b_{m} P^{m}) + Q
}
 which is clearly nilpotent (since $\{Q, b_{m}P^{m}\} = P^{2}$) 
 and has the properties
 $$
 Q_{new} b' = P^{2}\,, ~~~~ Q_{new} \b' = b' + b_{m}P^{m}\,, 
 ~~~~ Q_{new} c' = \g'\,, ~~~~~ Q_{new} \g' = 0\,. 
 $$
 This new BRST operator contains the old BRST charge $Q$ (which implements 
 the topological symmetry) but also the BRST symmetry of the diffeomerphisms with 
 the term $c' P^{2}$. The additional terms are needed  to render the BRST operator nilpotent. 
 The term $\gamma' b'$ is needed in order to generate a topological symmetry of the system 
 $c',b', \gamma', \beta'$ such that they do not enter the cohomology. 
 
In terms of $c', b', \gamma'$ and $\beta'$, 
we know what are the correct insertions 
for the moduli of a Riemann surface together with the 
supermoduli associated to $\gamma' $ and $\beta'$ 
(here we consider $g > 1$), namely one 
has to insert the following BRST invariant combinations
to reabsorb the zero modes of $b'$ and $\b'$ 
\eqn\gA{
\prod_{i}^{3(g-1) + N} \int \mu_{i} b'_{i} \{Q_{new}, \Theta( 
\int \mu_{i} \beta'_{i})\}  = 
\prod_{i}^{3(g-1) + N} \int \mu_{i} b'_{i} (b'_{i} + (b_{m} P^{m})_{i})
\delta( \beta'_{i}) 
} 
where $\mu_{i}$ are the Beltrami differentials. Since the 
fields $b'$ and $\beta'$ are decoupled, they can be integrated out leaving 
the insertion of $\prod_{i}^{3(g-1) + N} \int \mu_{i} (b_{m} P^{m})_{i}$ which 
are the insertions obtained in \gN. Notice that for genus 1, one has to 
reabsorb the isometries of the sphere and of the torus. This can 
be done by inserting the PCO $c' \delta(\gamma')$ to soak 
up the zero modes of $c'$ and $\gamma'$. For genus zero, we have 
already lengthly discussed it in the previous sections. 

The next step, we derive the insertion of the PCO $Z_{v_{i}}$. The number 
of $v_{i}$ is equal to the number of independent components of $w_{\a}$. 
We have to notice the following: there exist two solutions to the equation 
$\{Q, b^{\a} \} = \l^{\a} P^{2}$ (we refer to \BerkovitsPX\ for 
a complete analsys of this equation in the pure spinor framework). One 
solution is provided by $\l^{\a} b_{m} P^{m}$ and the second is 
given by $(\not\!P d)^{\a}$ (this is similar to the solution in 
\BerkovitsPX, in \AnguelovaPG\ and in \OdaBG) and the latter resembles the 
generator of the $\kappa$-symmetry of the Brink-Schwarz formulation 
of the superparticle.\foot{Recently paper \BerkovitsTW\ appeared, this provides a 
useful link between GS formalism and pure spinor formalism (as previously  
done so in \AisakaGA. It would be interesting to see the relation between PCO and 
the coupling to topological gravity for the critical superstring \GrassiPoli.
} Given those two solutions, 
we can form the BRST 
invariant combination 
$$
K^{\a} = (\not\!P d)^{\a} -  \l^{\a} b_{m}P^{m} \,.
$$
This combination is not only invariant, but it is also BRST-exact: 
\eqn\kA{
K^{\a} = \{Q, \not\!P w - b^{m} b^{n} \e_{mnp} (\g^{p}\l)\} \,.
}
This suggests that the operator $\Xi^{\a} = 
\not\!\!P w - b^{m} b^{n} \e_{mnp} (\g^{p}\l)$ can play 
the same role of $B= b_{m} P^{m}$ and 
its BRST transformed $P^{2}$ introduced above. 
In addition, we can check that the those operators form 
a closed algebra whose main commutation relation is 
given by 
\eqn\kB{
\{K^{\a}, K^{\b}\} = P^{m} \gamma^{\a\b}_{m} P^{2}\,, 
}
which is the usual relation between the $\kappa$-symmetry 
generators and the Virasoro constraint $P^{2} \sim 0$. As a 
check, one can also compute the commutation relation between 
$K^{\a}$ and $\Xi^{\b}$ to have 
\eqn\kC{
\{K^{\a}, \Xi^{\b} \} = - P^{m} \g^{\a\b}_{m} B\,,
}
which leads to \kB\ by acting with $Q$ on both side of the 
equal sign. 

Notice that the main difference between 
the operator $B$ and $\Xi^{\a}$ is the spinorial index carried 
by the second one. In addition, while $B$ is anticommuting, 
$\Xi^{\a}$ is commuting. This last observation suggests that 
we can introduce a new BRST operator for each spinorial 
component of $\Xi^{\a}$ and $K^{\a}$. It is however 
convenient to introduce again 
the gauge parameters $v_{(\a)}^{i}$ and define the 
BRST charge as follows
\eqn\cicA{
Q_{fin} = Q_{new} + \sum_{i}\Big[
\xi_{i}'' v^{i}_{\a} K^{\a} \delta'(v^{i}_{\a} \Xi^{\a}) + 
\gamma''_{i} 
\Big(\eta''^{i} + \delta(v^{i}_{\a} \Xi^{\a})\Big)
\Big]
}
where we converted the commuting operators $v^{i}_{\a} \Xi^{\a}$ 
into the fermionized ones $\delta(v^{i}_{\a} \Xi^{\a})$. 
Notice that we could have also used the other fermionized operators 
given by $\Theta(v^{i}_{\a} \Xi^{\a})$, but this is not a 
pseudoform. In addition, $\Theta(v^{i}_{\a} \Xi^{\a})$ does 
not carry any ghost number. On the other side, $\delta(v^{i}_{\a} \Xi^{\a})$ 
satisfies all the conditions.\foot{The BRST charge \cicA\ is 
very similar to the BRST operator $Q_{RNS} + \oint \eta$ proposed 
in 
\lref\BerkovitsUS{
N.~Berkovits,
JHEP {\bf 0108}, 026 (2001)
[arXiv:hep-th/0104247].
}
\BerkovitsUS\ where $\oint \eta$ is the BRST operator 
which reduces the large Hilbert space to the small one. 
} The action of the final BRST operator 
$Q_{fin}$ is 
given by 
\eqn\fA{
Q \beta''^{i} = \eta''^{i} + \delta(v^{i}_{\a} \Xi^{\a})\,, ~~~~
Q\eta''^{i} = v^{i}_{\a} K^{\a} \delta'(v^{i}_{\a} \Xi^{\a})
}
$$
Q\xi''_{i} = \gamma''_{i}\,, ~~~~~
Q\gamma''_{i} =0\,. 
$$
It is easy to show that the dependence 
upon the gauge parameters $v^{\a}_{i}$ is always BRST trivial. 

Following the previous derivation, we 
can derive the insertion to reabsorb the zero modes of 
$\eta''^{i}$ and of $\beta''^{i}$ 
\eqn\gB{
\prod^{n g + N}_{k=1} 
\oint_{A_{k}} \Big( \eta''^{k} v^{k}_{\a} K^{\a} \Big) 
\Big\{Q_{fin}, \Theta( \oint \b''^{k} )\Big\}
} 
Notice that at genus $g$ there are two zero modes for 
$\eta''^{i}$ and two zero modes $\beta''^{i}$. 
The integral is performed over an $A$ cycle of the 
Riemann surface (again here we dropped all this detail and we naively 
restrict ourselves to the worldline model). It is easy to check that those 
insertions are BRST closed, but not BRST exact. Notice that since 
the symmetry generated by $K^{\a}$ are not worldsheet symmetry, but 
rather target space symmetry, we expect that the 
insertions \gB\ are exaclty BRST closed (not up to  a total 
derivatives on the moduli). By computing 
the action of $Q_{fin}$ on $\beta''^{k}$ and using 
the fact that $\eta''^{i}$ is nilpotent (for each $i$) we 
get the following aswer
\eqn\gC{
\prod^{n g + N}_{k=1} 
\eta''^{k} v^{k}_{\a} K^{\a} 
\delta(\b''^{k}) \delta(v^{k}_{\a} \Xi^{\a})
} 
then finally integrating out $\eta''^{i}$ and $\beta''^{i}$, 
we obtain the insertion of the spacetime PCO
\eqn\gD{
\prod^{n g + N}_{k=1} Z_{v^{k}}.
}
where a suitable odd vector has been chosen which gives
$Z_{v^{k}} = \{Q, \Theta(v^{k}_{\a} \Xi^{\a})\}$. 
Notice that 
the dependence upon $v^{k}$ is entirely through BRST exact terms 
and they do not affect the amplitudes. 
 Finally, $ \Big\{Q_{fin}, \Theta( \oint \b''^{k} )\Big\}$ is the 
well-known worldsheet PCO constructed 
by fermonizing the superghost $\b''^{i}$ and it is interesting to 
see the interplay between a worldsheet PCO and the 
target space PCO.

\newsec{Future Directions}

The present analysis is a first step toward a more complete study of 
topological string theory on supertarget spaces. As we pointed 
out the theory of integration of superforms is directly related to 
the path integral formulation of superstrings and topological 
strings. Here we would like to indicate some future directions 
that we consider relevant for applications and developements. 

{\it i)} The analysis of boundary conditions of the sigma model 
given in sec. 2.2 and the study of A-branes and the analysis of their 
properties has be explored. 
As is well known the presence of fermionic degrees 
of freedom for the superstring D-branes drastically changes 
the spectrum of the theory. We suspect that also in the 
present context the study of A-branes might reveal some 
interesting properties. In addition, to our knowledge, the 
Super-Chern-Simon action is the only supersymmetric 
string field theory action constructed so far which is fully 
super-Poincar\'e invariant in the target space. We 
hope that the present construction can be useful to construct 
the full-fledged superstring field theory.

{\it ii)} Together with the A model, one can study the B model. In that 
case some anomalies coming from the holomorphic nature of the 
model should be properly compensated. It would be interesting 
to construct the top pseudoform to be integrated in the 
path integral. This certainly generalizes the usual form 
$\Omega$ of the super Calabi-Yau 
$$
\Omega = \e_{IJKL} Z^{I} dZ^{J} dZ^{K} dZ^{L} 
\e_{ABCD} d\Psi^{A} d\Psi^{B} d\Psi^{C} d\Psi^{D} 
$$ 
The observables of the model should be identified with 
superforms and the PCO should enter in the prescription 
for the amplitudes. 
It would be very interesting to understand the implications 
of these modifications in some specific calculation of amplitudes.

{\it iii)} The techniques of supergeometry can be applied to 
pure spinor string theory where the space of differential is 
modfied by the introduction of the pure spinor constraints. This 
will change the structure of the top form, but a 
complete analysis is missing. Interesting steps in that 
direction have been already achieved by Movshev and Schwarz 
in 
\lref\MovshevIB{
M.~Movshev and A.~Schwarz,
Nucl.\ Phys.\ B {\bf 681}, 324 (2004)
[arXiv:hep-th/0311132].
}
\lref\MovshevAW{
M.~Movshev and A.~Schwarz,
arXiv:hep-th/0404183.
}
\refs{\MovshevIB,
\MovshevAW}. In that context, one should be able to derive the 
multiloop prescription for pure spinor string theory by 
coupling it to topological gravity. 


\newsec{Acknowledgements}
We thank M. Porrati, N. Berkovits, P. Vanhove, 
L. Alvarez-Gaum\'e, W. Lerche, P. van Nieuwenhuizen, S. Theisen 
and W. Siegel for useful discussions. G. P.  thanks 
C.N. Yang Institute for Theoretical Physics at Stony Brook, where this 
work was started,  for the
hospitality.  This work was partly funded by NSF-grant PHY-0354776. 
G.P. is supported by the SFB 375 of DFG.

\newsec{Appendix: BV formalism} Here we view the BV formalism (see, cf. \as)
as an integral $I_{BV}$ of the BV differential form $\Omega_{BV}$
along the Lagrangian submanifold $\CL$ in the BV space:
\eqn\maineq{I_{BV}=\int_{\CL} \Omega_{BV}} The BV space $\CM$  is
equipped with the canonical odd symplectic form ${\omega}_{BV}$. One
can choose local coordinates to identify ${\CM}$ with ${\Pi}T^* N$
where $N$ is some (super)manifold, where the symplectic form has a
canonical form \eqn\bvform{ \omega_{BV}=\d Z_{a}^{+} \wedge \d Z^a }
where $Z^a$ denotes the (super)coordinates on $N$ and $Z_{a}^{+}$-
corresponding coordinates on the cotangent fiber.

The submanifold $\CL$ is Lagrangian with respect to the
canonical form $\omega_{BV}$ (in the physical literature its
generating function is called the gauge fermion). 

The BV differential form  $\Omega_{BV}$ is constructed out of two
ingredients \as: 
the BV action $S$ and the BV measure $\nu$: ${\Omega}_{BV} = \left(
{\nu} e^{- S} \right)$. 

The action $S$ must obey the so-called BV master equation:
\eqn\bvm{\{ S, S\}_{BV} := {\omega}_{BV}^{-1} ({\p}_l S \wedge {\p}_r
S) = 0}

One calls the coordinates $Z^a$ the fields and $Z^{+}_{a}$ the anti-fields.
Sometimes one distinguishes the classical part of $N$ and the
auxiliary fields used for gauge fixing.  
Also, the identification of the BV phase space
with ${\Pi} T^* N$ is not unique and is not global in general, so the
partition of all the fields involved  
on the fields and anti-fields is not unique.

The deformations of the action $S$ that preserve \bvm\ are (in the
first order approximation) the functions ${\Phi}$ on $\CM$ which
are $Q_{BV}$-closed, where the differential $Q_{BV}$ acts as
$Q_{BV} {\Phi} = \{ S , {\Phi} \}_{BV}$. The deformations which
are $Q_{BV}$-exact are trivial in the sense that they could be
removed by a symplectomorphism of ${\CM}$ (one has to make sure
that this symplectomorphism preserves $\nu$ to guarantee that the
quantum theory is not sensitive to such a $Q_{BV}$--exact term). 


\listrefs
\bye